\documentclass[%
 superscriptaddress,
 amsmath,amssymb,
 aps,
 pre,reprint,
]{revtex4-2}

\allowdisplaybreaks

\usepackage{graphicx}
\usepackage{appendix}
\usepackage{dcolumn}
\usepackage{bm}
\usepackage{xcolor}
\usepackage[normalem]{ulem}
\usepackage{xfrac}
\usepackage{hyperref}

\newcommand{\appropto}{\mathrel{\vcenter{
  \offinterlineskip\halign{\hfil$##$\cr
    \propto\cr\noalign{\kern2pt}\sim\cr\noalign{\kern-2pt}}}}}

\begin{document}

\preprint{APS/123-QED}

\title{\textbf{Condensation Transition in Entropy-Constrained Probability Spaces} 
}%

\author{Bautista Arenaza}
    \affiliation{Department of Applications of Physics and Biology to Health Sciences, Centro Atómico Bariloche, Argentina}
    \affiliation{Instituto Balseiro, Centro Atómico Bariloche, Argentina}
\author{Sebastián Risau-Gusman}
    \affiliation{Department of Applications of Physics and Biology to Health Sciences, Centro Atómico Bariloche, Argentina}
\author{Inés Samengo}
    \affiliation{Department of Applications of Physics and Biology to Health Sciences, Centro Atómico Bariloche, Argentina}
    \affiliation{Instituto Balseiro, Centro Atómico Bariloche, Argentina}
\author{Dami\'an G. Hern\'andez}\email{Contact author: damian.hernandez@ib.edu.ar}
    \affiliation{Department of Applications of Physics and Biology to Health Sciences, Centro Atómico Bariloche, Argentina}
    \affiliation{Instituto Balseiro, Centro Atómico Bariloche, Argentina}


\begin{abstract}
  Probability spaces are known to undergo a concentration of measure in high dimensions. In particular, if a $K$-dimensional probability distribution is sampled from a $(K-1)$-dimensional simplex with uniform measure, with high probability its empirical distribution of components is logarithmic.  Here, we extend this result to the subset of distributions populating level sets of de simplex defined by a fixed Shannon entropy $H_0$. We show that the empirical distribution of components of the large majority of sampled distributions undergoes a phase transition. A critical entropy $H_c \approx \log K - 1 + \gamma $ exists, below which the overwhelming majority of sampled distributions are found in a {\sl condensed state}, in which a single component captures a macroscopic fraction of the total probability,  while the remaining components form a homogeneous incompressible fluid background. We derive the magnitude of the critical entropy, as well as the empirical distribution of components both above and below threshold. These results provide insight into how the level sets of fixed entropy in the simplex are populated.  
  
\end{abstract}

\maketitle

\section{Introduction}

To understand organization of high-dimensional probability spaces, we here investigate the properties of the simplex $\Delta_{K-1}$, defined as the space of normalized $K$-dimensional probability vectors $\boldsymbol{\pi} = (\pi_1, \dots, \pi_K)$ with $\pi_i \ge 0$ and $\sum_{i=1}^K \pi_i = 1$. We focus on the large dimensional regime ($K \gg 1$), where concentration of measure strongly constrains the accessible regions of space \cite{talagrand1995concentration, raginsky2013concentration}. One of the consequences of this concentration is that, if a $K$-dimensional $\bm{\pi}$-vector is sampled from the simplex with uniform measure, its entropy $H = -\sum_j \pi_j \log \pi_j$ tends (in probability) to 
\begin{equation} \label{eq:hachetriangulito}
    H^\triangle = \ln K + \gamma - 1,
\end{equation} for $K \to +\infty$  \citep{Slud1978}, where $\gamma$ represents the Euler-Mascheroni constant. In the large-dimensional limit, the volume of the simplex around the level set of this particular entropy becomes exponentially dominant. Moreover, the ranked components of the distributions of the uniformly sampled simplex tend to cluster around a specific sequence, namely, \citep{Bairamov2010}
\begin{equation} \label{eq:triangulito}
    \pi^\triangle_j \to \frac{1}{K} \ \log \left(\frac{K}{j} \right).
\end{equation}

Our guiding question is: Are there similar regularities when considering only distributions of fixed entropy $H_0$? To address this matter, in this paper we consider the simplex endowed with the uniform measure and examine the level sets of constant Shannon entropy, $H(\boldsymbol{\pi}) = -\sum_{i=1}^K \pi_i \log \pi_i = H_0$.

\begin{figure}[ht]
\begin{center}
    \includegraphics[scale = 0.43]{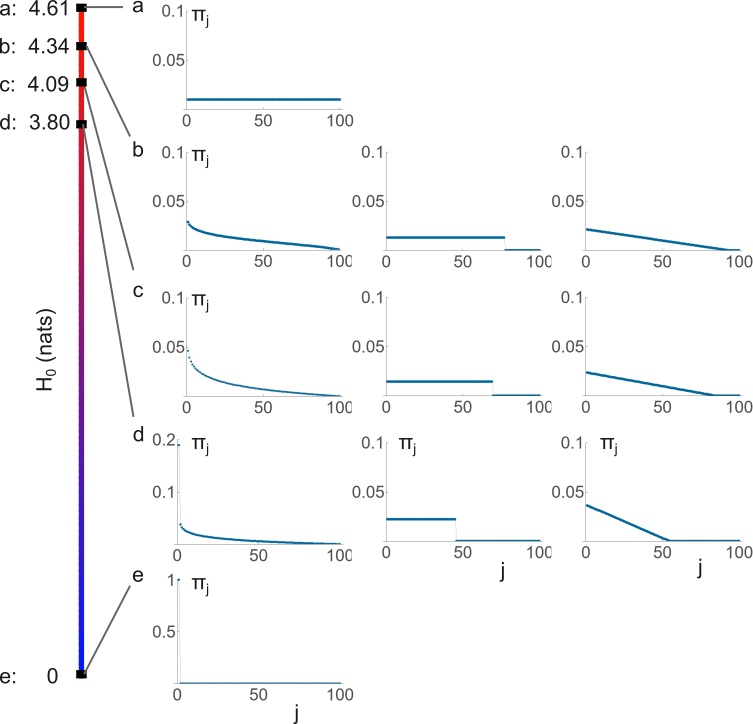}
\end{center}
\caption{\label{fig:samplesmicro} Components $\pi_j$ (sorted in decreasing order) of sampled probability vectors $\bm{\pi}$, belonging to a level set of fixed $H_0 = -\sum_j \pi_j \log \pi_j$, for $K = 100$. Panels in the left column:  $\bm{\pi}$-vectors most frequently observed when sampling with uniform measure on the level set. Middle and right panels: $\bm{\pi}$-vectors belonging to the same level sets, but virtually never sampled. Coloured line on the left: Entropy $H_0$ fixed to a different value in each row \emph{a, b, c, d, e} (linear scale).}
\end{figure}

Figure~\ref{fig:samplesmicro} 
shows examples of $K  = 100$-dimensional probability vectors $\bm{\pi}$, with their components $\pi_j$ ranked in decreasing order, sampled from different level sets of entropy. 

The level set $H_0 = \log (K)\approx 4.61$ nats contains a single $\bm{\pi}$ vector, namely, the uniform distribution, in which all components are exactly equal and take the value $\pi_j = \sfrac{1}{K}$ (top row, case $a$). No other $\bm{\pi}$ vector can be sampled from this level set. As the entropy $H_0$ diminishes (subsequent rows in Fig.~\ref{fig:samplesmicro}) more $\bm{\pi}$ vectors fulfil the entropy constraint. A few examples are shown in the three columns corresponding to each row $b, c$ or $d$. All the $\bm{\pi}$ vectors in the same row have the same entropy $H_0$, so in principle, any of them can be picked by sampling the corresponding level set. Yet, as the dimension $K$ grows, the vast majority of the actually sampled vectors have ranked components that look like those in the left column. Examples with piecewise constant (middle column) or linearly decreasing (right) components are virtually never observed. At a critical value of the entropy $H_0$, which in Fig.~\ref{fig:samplesmicro} corresponds to case $c$, the ranked components of the actually sampled $\bm{\pi}$-vectors are close to the law given by Eq.~\eqref{eq:triangulito}, for vectors sampled from anywhere in the simplex. For $H_0$ values below the critical entropy, most sampled $\bm{\pi}$-vectors exhibit a condensation phenomenon: a single component carries a finite fraction of the total probability ($\pi_1 \approx 0.2$ in the left panel of row $d$ in Fig.~\ref{fig:samplesmicro}), while the remaining components form a continuous background (note this panel has a different scale on the $y$ axis). Throughout the paper, we call the single large component, of order unity, the ``\emph{condensate}''. The remaining small components, of order $\sfrac{1}{K}$, are here referred to as ``\emph{fluid}''. The size of the condensed component increases as $H_0$ diminishes.
For all $H_0 < H_c$, the ranking of the fluid components is still well described by Eq.~\eqref{eq:triangulito}, except for a global scaling factor that depends on the size of the condensed component to ensure normalisation. Decreasing  $H_0$ all the way down to $H_0 = 0$ yields $\bm{\pi}$-vectors that represent deterministic distributions, in which one component takes unit value, and the remaining ones vanish (panel $e$ in Fig.~\ref{fig:samplesmicro}).

The goal of this paper is to explain these phenomena. We derive the value of the critical entropy, and clarify why the condensation takes place. We also derive the value of the condensed component, and the rank of the fluid components actually sampled both above and below the critical threshold, in the limit of large $K$.  




\section{Microcanonical and canonical descriptions of an ensemble with constrained entropy} 

\label{sect:sna}

To investigate the properties of the level sets of entropy $H(\bm{\pi}) = H_0$ in high-dimensional simplices, we define the partition function
\begin{align} 
    Z(H_0) &= \int_0^1 {\rm d}\pi_1 \dots \int_0^1 {\rm d}\pi_K \ \delta\left(\sum_{i=1}^K \pi_i - 1\right) \nonumber \\
    &\quad \times \delta\big(H(\boldsymbol\pi) - H_0\big), \label{eq:partition_function}
\end{align}
the value of which equals the volume of the level set under the uniform measure. Representing both Dirac deltas in exponential form,
\begin{equation}
    \delta(x) = \frac{1}{2\pi i} \int_{-i\infty}^{+i\infty} {\rm d}s \ \mathrm{e}^{-sx}, \label{eq:deltaintegral}
\end{equation}
decouples the two constraints, so that
\begin{equation} \label{eq:sym_Z}
    Z(H_0) \propto \mathrm{e}^{s + tH_0}\int_{-i\infty}^{+i\infty} \mathrm{d}s \int_{-i\infty}^{+i\infty} \mathrm{d}t \ z(s, t)^K,
\end{equation}
where 
\begin{equation}
    z(s,t) = \int_0^1 {\rm d}\pi \ \mathrm{e}^{-s \pi + t \pi \log \pi}. \label{eq:zetachica}
\end{equation}

\subsection{The saddle-point approximation}

The standard procedure to get an analytical approximation to $Z(H_0)$ \citep{SzavitsNossan2014PRL, SzavitsNossan2014JPA} is to use a saddle-point approximation, by which the integrand in Eq.~\eqref{eq:sym_Z} is assumed to be sharply peaked at a single pair $(s, t)$, the value of which depends on $H_0$ and $K$, and is defined by the saddle-point equations
\begin{equation} \label{eq:saddle-point}
  \left\{
  \begin{aligned}
    &\partial_s \log z(s,t) = -\sfrac{1}{K} \\ 
    &\partial_t \log z(s,t) = -\sfrac{H_0}{K}.
  \end{aligned}
  \right.
\end{equation}
As a result, 
\begin{align}
    Z(H_0) \appropto \mathrm{e}^{s + tH_0} \ z(s, t)^K. \label{eq:zaprox}
\end{align}
The factor $z(s, t)^K$ is the partition function of a different system, namely, one composed of $K$ independent and identically distributed variables drawn from the single-site density
\begin{equation} \label{eq:single_site_dist}
  f(\pi) = \frac{\mathrm{e}^{-s\pi+t\pi\log\pi}}{z(s,t)} .
\end{equation}

Understanding the equivalence between the original system and the effective one requires a closer examination.
In Eq.~\ref{eq:partition_function}, the normalization and entropy constraints were imposed exactly, implying that the components of $\bm{\pi}$ were not independent. The partition function $Z(H_0)$ defined a \emph{microcanonical} ensemble over the probability simplex. By expressing the Dirac delta functions in their integral representation (Eq.~\ref{eq:deltaintegral}), and after assuming a single saddle dominates the integral, the microcanonical system became asymptotically equivalent to that of a \emph{canonical} ensemble. 

The Lagrange multipliers $s$ and $t$ introduced through the integral representation of the constraints are conjugate to the normalization and entropy restrictions, respectively, and their value is determined by the saddle-point equations \eqref{eq:saddle-point}. In the canonical formulation, the components of $\bm{\pi}$ become independent, and drawn from the single-site distribution $f(\pi)$. Defining the quantities
\begin{equation} \label{eq:emeyhache}
    \begin{split}
    M &= \sum_{i = 1}^K \pi_i \\
    H &= -\sum_{i = 1}^K \pi_i  \log \pi_i,
    \end{split}
\end{equation}
in the canonical (or single-site) formulation, the saddle-point conditions of Eqs.~\eqref{eq:saddle-point} become
\begin{equation}
\begin{split}
\left\langle M \right\rangle_f &= K \langle \pi \rangle_f =1, \\
\left\langle H \right\rangle_f &= - K \langle \pi \log \pi \rangle_f =H_0,
\end{split} \label{eq:const} 
\end{equation}
where $\langle \cdot \rangle_f$ represents the expected value weighted by the density $f$ of Eq.~\eqref{eq:single_site_dist}.
Consequently, the constraints on the normalization and the entropy are no longer enforced for every $\bm{\pi}$ vector, but only in expectation. The quantities $M$ and $H$ now fluctuate around $1$ and $H_0$, respectively, while their expectation values remain fixed. In other words, realizations of the canonical ensemble are allowed to lie outside of the level curve of entropy $H_0$, and even outside the simplex. As we show in Sect.~\ref{sec:validitysn}, the validity of the saddle-point approximation yielding Eq.~\eqref{eq:zaprox} depends on whether the fluctuations outside of the simplex and outside the level set of entropy tend to vanish in the large $K$ limit. This may or may not happen depending on whether $H_0$ is above or below a critical entropy (see below).

\subsection{Behaviour of the single-site density}

The shape of $f(\pi)$ depends on the values of $s$ and $t$ that solve the saddle-point equations, which in turn are determined by $K$ and $H_0$.

The parameter $s$ is conjugate to the normalisation condition. The first saddle-point equation in \eqref{eq:const} implies that $\langle \pi \rangle_f = \sfrac{1}{K}$ which, in the large $K$ limit, becomes $\langle \pi \rangle_f \to 0$. This condition, when applied to Eq.~\eqref{eq:single_site_dist}, forces $s$ to be strictly positive and increasing with $K$, for any $t$ value. 

The parameter $t$ is conjugate to the Shannon entropy \cite{amari2000methods} and controls the homogeneity of the components $\pi_i$ of the $K$-dimensional probability vector.  For a fixed $K$, the constraints can be used to show that $H_0$ is a strictly decreasing function of $t$. 

\begin{figure}[t]
  \centering
  \includegraphics[width=0.42\textwidth]{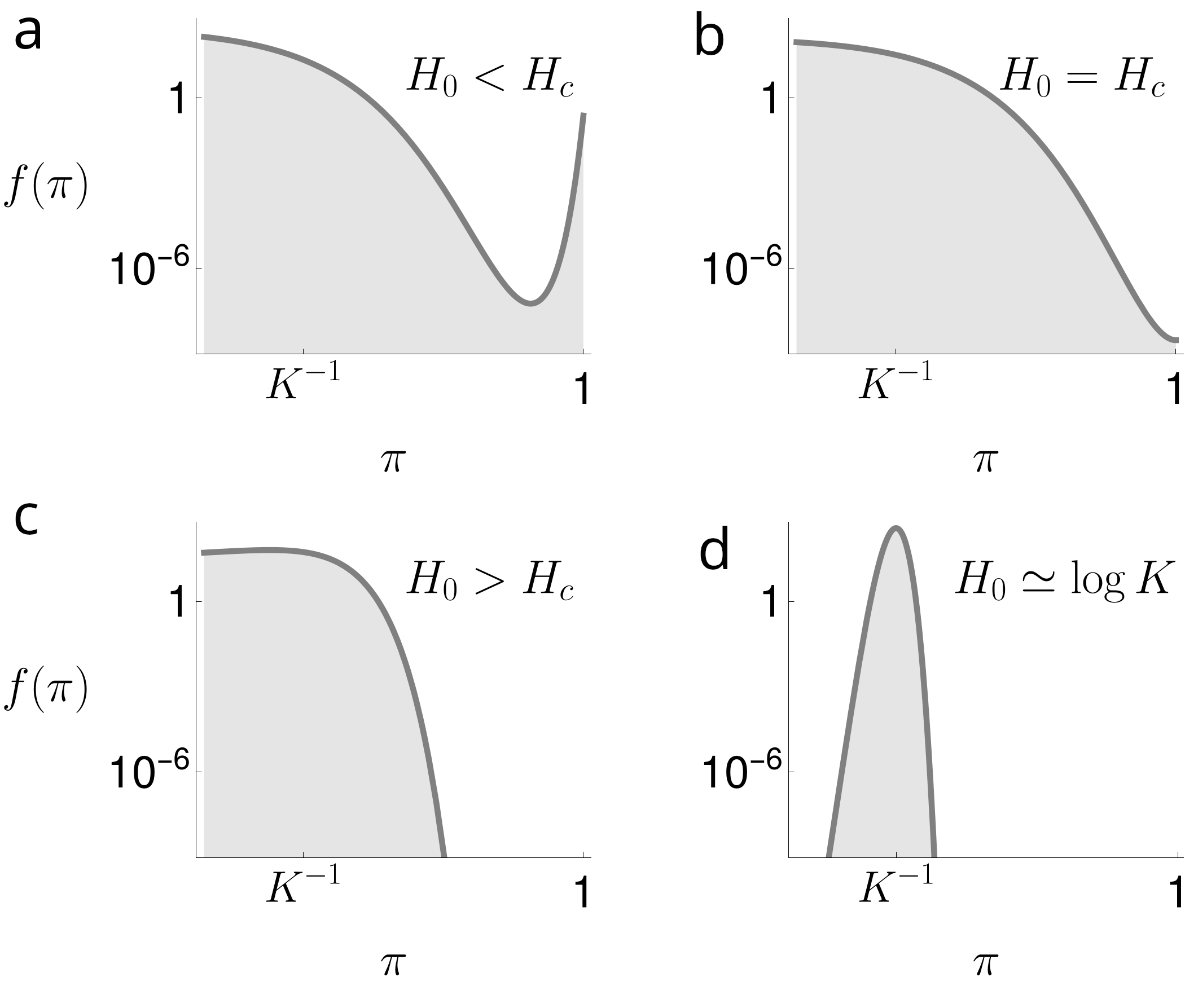}
  \caption{Single-site density $f(\pi)$ for four different values of the entropy $H_0$, diminishing from \emph{a} to \emph{d}. Both axes are in logarithmic scale. For  $H_0$ close to its maximal value $\log K$, (panel $d$) $f(\pi)$ is sharply peaked around $\sfrac{1}{K}$. As $H_0$ decreases ($c$), the peak of the distribution becomes asymmetric and the maximum shifts to the left. For $H_0 = H_c$ ($b$), a local minimum appears at $\pi_c = 1$. For even lower entropies ($a$), $\pi_c$ shifts to the left, resulting in a macroscopic peak at $\pi=1$. This peak signals the condensation, assigning a finite fraction of the total probability mass to $\pi$ values close to unity.}
  \label{fig:density}
\end{figure}

The single-site density $f(\pi)$ is illustrated  in Fig.~\ref{fig:density}, with $H_0$ increasing from \emph{a} to \emph{d}. Setting the derivative of the exponent of Eq.~\eqref{eq:single_site_dist} to zero reveals an extremum located at $\pi_c = \exp(s/t - 1)$, which may or may not fall within the physically meaningful interval $\pi \in [0, 1]$ depending on the values of $s$ and $t$.

For large entropy values, i. e. $H_0 \lesssim \log K$, the parameter $t$ is large and negative. In this regime, $f(\pi)$ is concave, and close to a delta-distribution peaked at $\pi_c = \sfrac{1}{K}$  (panel $d$ in Fig.~\ref{fig:density}). As $H_0$ diminishes, $t$ grows, $f(\pi)$ widens, and the maximum at $\pi_c$ shifts to the left. When $t = 0$, $\pi_c$ vanishes, and Eq.~\eqref{eq:single_site_dist} switches from concave to convex, since at that point, the second derivative of Eq.~\eqref{eq:single_site_dist} also vanishes at $\pi_c$. If $H_0$ diminishes further, $t$ is increased beyond zero, and $f(\pi)$ becomes a monotonically decreasing function in the interval $[0,1]$ (panel $c$ in Fig.~\ref{fig:density}). The extremum $\pi_c = \exp(s/t - 1)$ is now a minimum located at a physically meaningless value $\pi_c > 1$. This minimum shifts to the left as $t$ grows, and touches the point $\pi = 1$ when $t$ becomes equal to $s$ (panel $b$ in Fig.~\ref{fig:density}). 
When the entropy is lowered even further, $\pi_c$ shifts to the left (panel $a$ in Fig.~\ref{fig:density}), and the density $f(\pi)$ becomes bimodal. Most of its mass is concentrated at $\pi \lesssim \sfrac{1}{K}$. Yet, a mass of order $\sfrac{1}{K}$ is located around a second peak at $\pi = 1$. Therefore, if $K$ independent samples are drawn from $f(\pi)$, one or a few samples may well have values of order unity. This second peak underlies the phenomenon of condensation.

\subsection{The ranked components of most of the vectors belonging to the level set}

The single-site distribution $f(\pi)$ of Eq.~\eqref{eq:single_site_dist} governing the behaviour of each component of the canonical ensemble emerged after the saddle-point approximation. Yet, that same distribution can also arise without invoking the approximation. It can also be obtained by searching for the density $f(\pi)$ that maximises the differential entropy $h(f) = -\int f(\pi) \log f(\pi) \mathrm{d}\pi$ under the constraints $\langle \pi \rangle_f = \sfrac{1}{K}$ and $-\langle \pi \log \pi \rangle _f = \sfrac{H_0}{K}$. This is no coincidence. 

A simple counting argument, presented in Appendix~\ref{sec:pelotitas}, shows that the volume of the simplex occupied by $\bm{\pi}$ vectors with a specific empirical distribution of components $\hat{f}(\pi | \bm{\pi}) = \sum_i \delta(\pi - \pi_i) / K$ is proportional to $\mathrm{e}^{K h(\hat{f})}$. Therefore, finding the $\hat{f}(\pi)$ that maximises the differential entropy subject to the given constraints amounts to finding the distribution of components that becomes asymptotically dominant in the chosen level set of the microcanonical system. The single-site density $f(\pi)$ is the solution to that maximisation problem (Appendix~\ref{sec:pelotitas}).

This result has a practical implication: It enables the derivation of the most probably sampled sequence of ranked components, examples of which were shown in the left column of Fig.~\ref{fig:samplesmicro}. The number of components $j(\pi)$ of a $\bm{\pi}$ vector sampled from a level set of entropy $H_0$ that lie above a certain value $\pi$ is proportional to 
\begin{equation} \label{eq:rank}
    j(\pi) = K \left[ 1 - F(\pi)\right] = K \int_\pi^1 f(\pi') \ \mathrm{d}\pi'
\end{equation}
where $F(\pi)$ is the cumulative distribution associated with $f(\pi)$.
In order to plot this ranking, values of $s$ and $t$ must be obtained by solving the saddle-point equations \eqref{eq:saddle-point}, which is typically done numerically. Then, the integral in Eq.~\eqref{eq:rank} can also be solved numerically, yielding a relation between the value of $\pi$ and the order $j$ in the rank.

\subsection{Deriving the critical threshold}

The critical threshold $H_c$ is the entropy at which a local minimum appears at $\pi_c = 1$ in the single-site distribution $f(\pi)$ (panel $b$ in Fig.~\ref{fig:density}).  For $H_0 < H_c$ (panel $a$), $f(\pi)$ becomes bimodal. The threshold marks the minimum entropy achievable by only fluid components, without macroscopic clumping. Below the threshold, as shown in the next section, condensation forces a non-infinitesimal fraction of $M = \sum_i \pi_i$ to concentrate into a single dominant component: the condensate.

The condition $\pi_c = \mathrm{e}^{s / t - 1} = 1$ implies $s = t$.
To calculate the value of the critical entropy $H_c$, in Appendix~\ref{sec:single_site_int}, we derive an analytical approximation for $z(s, t)$ (Eq.~\ref{app:eq:logz_approx}), which tends to be exact in the limit of large $K$. In Appendix~\ref{sec:condensation_threshold},  this approximation is used to evaluate the saddle-point equations \eqref{eq:saddle-point} along the diagonal $t = c$. Neglecting terms beyond order $\mathcal{O}(1/\log K)$, we obtain 
\begin{equation} \label{eq:Hc}
  H_c \approx \log K + \gamma - 1 - \frac{\pi^2-9}{3\log K}.
\end{equation}
This equation differs from the typical entropy $H^\triangle$ obtained by sampling the whole simplex with uniform measure (Eq.~\ref{eq:hachetriangulito}) in only the last term, of order $\mathcal{O}(1/\log K)$.
Therefore, the critical entropy $H_c$ is only slightly below the level set that occupies the largest fraction of the entire simplex.

\section{Analysis of the validity of the saddle-point approximation}

\label{sec:validitysn}

We here analyse the validity of the saddle-point approximation through the tilting formulation  \cite{goutis01081999}, which we now apply to the partition function $Z(H_0)$ of Eq.~\eqref{eq:partition_function}.
On the support of the two Dirac deltas, the relation
\begin{equation*}
    \sum_{i=1}^K \big( -s\pi_i + t\,\pi_i\log\pi_i \big) = -s - tH_0
\end{equation*}
holds pointwise for any $(s,t)$.
Therefore
\begin{align}
  Z(H_0)
  =& \int_0^1 {\rm d}\pi_1 \dots \int_0^1 {\rm d}\pi_K \ \delta\left(\sum_{i=1}^K \pi_i - 1\right) \delta\left(H(\boldsymbol \pi) - H_0\right) \nonumber \\
  =& z(s,t)^K \int_0^1 {\rm d}\pi_1 \dots \int_0^1 {\rm d}\pi_K \ \delta\left(\sum_{i=1}^K \pi_i - 1\right) \times \nonumber \\
  & \times  \delta\left(H(\boldsymbol \pi) - H_0\right) \prod_{i=1}^K f(\pi_i) \nonumber \\
  =& \mathrm{e}^{s + tH_0} \ z(s,t)^K\ \tilde p\big(M = 1,\ H = H_0\big), \label{app:eq:saddle-point_app}
\end{align}
where $\tilde p$ is the joint density of the pair $(M, H)$ defined in Eq.~\eqref{eq:emeyhache} when the $K$ components are drawn independently from $f(\pi)$,
\begin{align}
  \tilde{p}(M, H) =& \int_0^1 {\rm d}\pi_1 \dots \int_0^1 {\rm d}\pi_K \ \delta\left(\sum_{i=1}^K \pi_i - M\right) \times \nonumber \\
  & \times \delta\left(H(\boldsymbol \pi) - H\right) \prod_{i=1}^K f(\pi_i). \label{eq:pmonio}
\end{align}
The saddle-point equations~\eqref{eq:const} make $(1, H_0)$ the mean of $\tilde p$.
The approximation consists of evaluating $\tilde p$ at its mean using the Central Limit Theorem, that is,
\begin{equation} \label{eq:gaussian_step}
  \tilde p\big(1, H_0\big) \approx \frac{1}{2\pi\sqrt{\det C}},
\end{equation}
where $C$ is the $2 \times 2$ covariance matrix of the pair of variables $(M, H)$ under the measure $\tilde p$. In Appendix \ref{sec:failure}, we show that the Gaussian assumption is only valid when $H_0 \ge H_c$. 

To understand why the approximation may fail, we note that in Eqs.~\eqref{eq:emeyhache}, both $M$ and $H$ are expressed as sums of samples which, in Eq.~\eqref{eq:pmonio}, are governed by the product $\prod_i f(\pi_i)$ describing independent variables. At first sight, this structure suggests that the Central Limit Theorem should guarantee convergence to a Gaussian distribution in the large-$K$ limit. However, the density $f(\pi)$ from which each $\pi_j$ is sampled changes with $K$, because both $s$ and $t$ are themselves $K$-dependent. The applicability of the Central Limit Theorem therefore hinges on how $f(\pi)$ evolves with $K$. As $K$ grows, $f(\pi)$ becomes increasingly concentrated around its single maximum (for $H_0 \ge H_c$) or its two maxima (for $H_0 < H_c$). Appendix~\ref{sec:failure} proves that, as a consequence, the Gaussian limit is valid only above the critical entropy.



For $H_0 < H_c$, Eqs.~\eqref{eq:sym_Z} and \eqref{eq:zetachica} still hold. However, it is no longer true that the distribution of $(M, H)$
is asymptotically Gaussian around the values prescribed by the constraints. Below the threshold, $(M, H)$ is governed by a weighted mixture of narrow bivariate Gaussian peaks, the realizations of which satisfy the normalization constraint on average. In this section we sketch the conceptual base of this result, and Appendix \ref{sec:failure} provides the explicit mathematical derivation.

The mixture structure follows from the bimodal shape of the single-site distribution $f(\pi)$ for $H_0<H_c$ (Fig.~\ref{fig:density}, left panel). With non-negligible probability, one component---the condensate---takes values near the peak of $f(\pi)$ close to unity, while the remaining $K-1$ fluid components are concentrated near the peak at the origin. Let $w_1$ denote the probability mass of the condensed mode, defined as the integral of $f(\pi)$ from $\pi_c$ to $\pi=1$. Since $\langle\pi\rangle_f=1/K$, this weight must scale as $1/K$, so we write $w_1=c/K$, with $0\le c\lesssim 1$.

Consider now the full vector of $K$ independent components drawn from $f(\pi)$. The number $N$ of components sampled from the condensed mode is governed by a binomial distribution,
\[
N\sim\mathrm{Binomial}(K,c/K),
\]
which converges to a Poisson distribution of mean $c$ in the limit of large $K$. Each condensed component has magnitude $c$ (which is of order unity) and contributes asymptotically no entropy, since $-\pi\log\pi\rightarrow0$ as $\pi\rightarrow1$. The remaining components form the fluid background and, by the normalization condition, sum up to approximately $1-c$, with Gaussian fluctuations of order $K^{-1/2}$ around this value. Consequently, the total sum $M$ of the components of the vector $\bm{\pi}$  converges in distribution to
\begin{equation}
M = \sum_i\pi_i\longrightarrow (1-c)+N,
\qquad N\sim\mathrm{Poisson}(c),
\label{eq:mass_limit_law}
\end{equation}
where the Gaussian fluctuations of the fluid contribution were neglected. The probability density governing $M$ is therefore a sequence of narrow Gaussian peaks centred at $1-c$, $2-c$, $3-c,\dots$, with Poisson weights determined by the number of condensed components $N$. The mean of Eq.~\eqref{eq:mass_limit_law} is $(1-c)+c=1$, consistent with the saddle-point equations. However, for $0<c<1$, none of the peaks is centred exactly at $1$, implying that most of the $\bm{\pi}$ vectors sampled from the equivalent canonical system fall outside of the simplex. They rather form clusters in $\mathbb{R}^K$ that are concentrated on hyperplanes parallel to the simplex, corresponding to different values of $M$. The cluster with no condensed components is centred around $M=1-c$, the one with a single condensed component around $M=2-c$, and so forth.

The failure of the symmetric saddle-point treatment becomes now evident: the value of the constraint $\langle M \rangle = 1$ lies in the exponentially suppressed region between $N = 0$ and $N = 1$, rather than at the maximum of a Gaussian distribution. Consequently, the evaluation of the integrand at the saddle-point is a very poor approximation to the value of the integral, and the situation worsens as $K$ increases.

\section{An asymmetric saddle-point approximation}

The Gaussian assumption is the only approximation in the derivation of the partition function in Sect.~\ref{sect:sna}, so the failure for $H_0 < H_c$ must stem from the assumption ceasing to hold. Here we derive an alternative strategy based on an asymmetric saddle-point treatment, where the condensate component is separated out, and treated differently from the remaining fluid components. The fluid components are then  renormalised to define an element of a simplex of one lower dimension, for which the saddle-point approximation is once again valid.  

Let $\alpha$ denote the maximal component of $\boldsymbol \pi$. We define  $\boldsymbol{\tilde \pi} = (\tilde{\pi}_1, \dots, \tilde{\pi}_{K-1})$ as the vector obtained by removing $\alpha$ from
$\bm{\pi}$ and rescaling the remaining components by $1/(1-\alpha)$, so they sum to unity. Separating out the $\alpha$ component in the partition function of Eq.~\eqref{eq:partition_function},
\begin{align*}
  Z(H_0)
  &= K \int_0^1 \mathrm{d}\alpha \int_0^\frac{\alpha}{1-\alpha} \mathrm{d}\tilde{\pi}_1 \dots \int_0^\frac{\alpha}{1-\alpha} \mathrm{d}\tilde{\pi}_{K-1} \\
  &\quad \times (1-\alpha)^{K-2} \delta\left(\sum_{i=1}^{K-1} \tilde{\pi}_i - 1\right) \\
  &\quad \times \delta\left[H(\alpha, 1-\alpha) + (1-\alpha) H(\boldsymbol{\tilde \pi}) - H_0\right].
\end{align*}
The nesting property of the entropy states that 
\begin{equation}
    H(\boldsymbol \pi) = H(\alpha,1-\alpha) + (1-\alpha) H(\boldsymbol{\tilde \pi}), \label{eq:nesting}
\end{equation}
implying that the total entropy decomposes into a contribution from distinguishing the condensate from the fluid, and a contribution from the distribution of probability among the fluid components. Using the exponential representation of the Dirac delta to enforce the constraints, 
\begin{align} \label{eq:asym_Z}
  Z(H_0)
  &\propto \int_{-i\infty}^{+i\infty} \mathrm{d}s \int_{-i\infty}^{+i\infty} \mathrm{d}t \int_0^1 \mathrm{d}\alpha \ \mathrm{e}^{-G(s,t,\alpha,H_0)}.
\end{align}
The exponent function $G(s, t, \alpha, H_0)$ is defined as
\begin{align*}
  G(s,t,\alpha,H_0)
  &= -s-t H_0 +t H(\alpha,1-\alpha) \\
  &\quad - (K-2) \log(1-\alpha) \\
  &\quad -(K-1) \log \tilde{z}(s,t,\alpha)
\end{align*}
and the sub-component single-site integral is
\begin{equation}
  \tilde{z}(s,t,\alpha) = \int_0^\frac{\alpha}{1-\alpha} \mathrm{d}\tilde\pi \ \mathrm{e}^{-s\tilde\pi+t(1-\alpha)\tilde\pi\log\tilde\pi}. \label{eq:zetatilde}
\end{equation}
For each value of $\alpha$, Eqs.~\eqref{eq:asym_Z} and \eqref{eq:zetatilde} define an equivalent canonical system for $K - 1$ independently sampled fluid components, with marginal density
\begin{equation}
        \tilde{f}(\tilde{\pi}) = \frac{\mathrm{e}^{-s\tilde\pi+t(1-\alpha)\tilde\pi\log\tilde\pi}}{\tilde{z}(s,t,\alpha)}, \ \ \ \mathrm{for} \ \tilde{\pi} \in \left[0, \frac{\alpha}{1 - \alpha}\right]. \label{eq:efedepimonio}
\end{equation}
In this context, the saddle-point approximation applied to \eqref{eq:asym_Z} is
\begin{align*}
  Z(H_0) \appropto \mathrm{e}^{-G(s,t,\alpha,H_0)},
\end{align*}
where $(s,t,\alpha)$ are the solutions to the saddle-point equations $\partial_s G = \partial_t G = \partial_\alpha G = 0$. 

The symmetric saddle-point approximation lost validity for $H_0 < H_c$ because the bimodal nature of $f(\pi)$ implied that $M$ and $H$ were no longer governed by a Gaussian distribution. In the asymmetric formulation, separating out the condensate component yields a single-site distribution $\tilde{f}(\tilde{\pi})$ for the fluid components (Eq.~\ref{eq:efedepimonio}) that is monotonically decreasing over the interval $\tilde{\pi} \in [0,\alpha/(1-\alpha)]$. This property is true for every triplet $(s, t, \alpha)$ that solves the saddle-point equations, as can be verified numerically. Consequently, the analysis of Sect.~\ref{sec:failure} applies with $c=0$, implying that the random variables $\tilde{M}$ and $\tilde{H}$ (defined as in Eq.~\ref{eq:emeyhache}, but in terms of the $\tilde{\pi}_j$) are jointly Gaussian, with means given by their constrained values.

The condensed regime can now be studied by simultaneously tracking the behaviour of the largest component $\alpha$ and the sub-component entropy $H(\boldsymbol{\tilde \pi})$. Note that in the present formalism, the value $\alpha$ of the condensate component is not part of the equivalent canonical system. When fixing $H_0$, the saddle point equations fix the value of $\alpha$, so that only the fluid components are sampled independently from each other in the canonical equivalent system.  

\subsection{The entropy of the normalised fluid components}

\begin{figure}[h]
  \centering
  \includegraphics[width=0.45\textwidth]{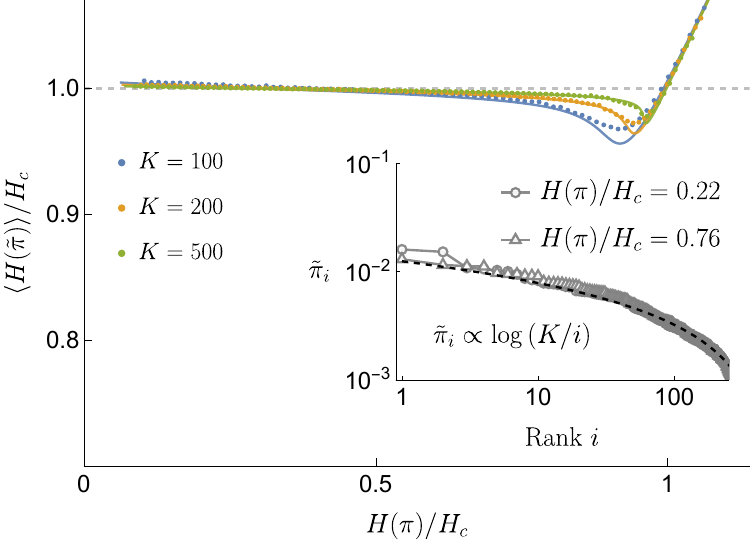}
  \caption{Entropy and ranking of the renormalised fluid components of $\bm{\pi}$-vectors sampled uniformly from the level set $H(\bm{\pi}) = H_0$. The sampling procedure is detailed in Appendix~\ref{sec:simul}.  Coloured dots: Scaled entropy $H(\tilde{\bm{\pi}})/ H_c$ for different dimensions $K$. The vertical axis starts at $0.7$, so the departure of $H(\tilde{\bm{\pi}})$ from $H_c$ is small, and tends to vanish as $K$ increases.  Full coloured lines: Entropy of the normalised, fluid components obtained by plugging the solution to Eq.~\eqref{app:eq:saddle-point} into Eq.~\eqref{app:eq:meanHx}. \emph{Inset}: Ranked fluid components of individual $\bm{\pi}$-vectors sampled uniformly from the level sets of two different values of $H_0$. Dotted line: ranking $\pi^\triangle(j)$ of Eq.~\eqref{eq:triangulito}, valid for uniformly sampling the simplex. 
  \label{fig:h_samples}}
\end{figure}

A remarkable fact is that, as shown in Appendix \ref{sec:app:mean}, for $H_0 < H_c$ and large $K$, the entropy of the normalised fluid components $\tilde{\bm{\pi}}$ essentially remains fixed at $H_c$, almost irrespective of $H_0$. More precisely, up to $\mathcal{O}(1/\log K)$,
\begin{equation}
    \left\langle H(\boldsymbol{\tilde \pi}) \mid H(\boldsymbol{\pi}) =  H_0 \right\rangle \approx \log(K-1) + \gamma - 1 -  \frac{\pi^2 - 9}{3 \log K}. \label{eq:entropymonio}
\end{equation}
Since $\log (K-1) = \log K + \mathcal O(1/K)$, and $\mathcal O(1/K) < \mathcal O(1/\log K)$, $\left\langle H(\boldsymbol{\tilde \pi}) \mid H(\boldsymbol{\pi}) =  H_0 \right\rangle$ coincides with $H_c$ up to $\mathcal O(1/\log K)$. In Appendix \ref{sec:variance}, we also show that $\mathrm{Var}(H(\boldsymbol{\tilde \pi}) | H(\boldsymbol{\pi}) = H_0) = \mathcal{O}(1/K)$, implying that the distribution of entropies of the (normalised) fluid components concentrates around the critical entropy $H_c$, for all $H_0 < H_c$.

The entropy of the normalised fluid components is displayed in Fig.~\ref{fig:h_samples} for $\bm{\pi}$-vectors sampled uniformly from the level set $H(\bm{\pi}) = H_0$. The sampling procedure is detailed in Appendix \ref{sec:simul}.  The sampled entropies are contrasted with the analytical approximation $\left\langle H(\boldsymbol{\tilde \pi}) | H(\boldsymbol{\pi}) \right \rangle \approx H_c$ (dashed line), and the numerical solution of Eq.~\eqref{app:eq:saddle-point} plugged into \eqref{app:eq:meanHx} (full coloured lines). For $K = 100$ (blue), the samples (dots) are well described by the theory (solid line), and for $K = 500$ (green), the match is excellent.

\subsection{Deriving the size of the condensate component}

If the entropy $H(\bm{\tilde\pi})$ of the fluid components is essentially fixed at $H_c$, by conveniently choosing the value of $\alpha$, Eq.~\eqref{eq:nesting} can reach any entropy $H(\boldsymbol\pi) = H_0$. In this section we discuss the resulting relation between $\alpha$ and $H_0$.

\begin{figure}[h]
  \centering
  \includegraphics[width=0.48\textwidth]{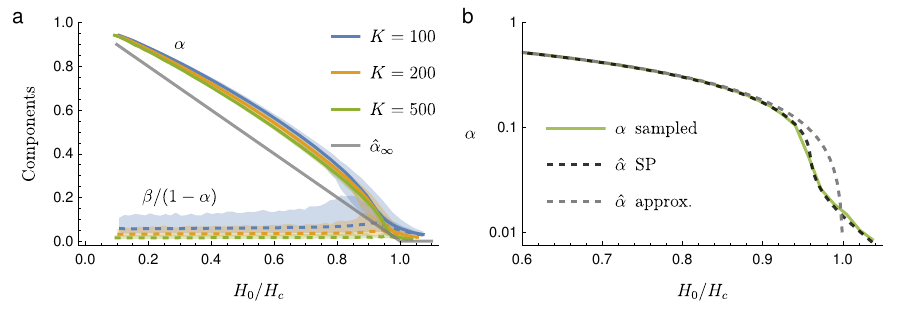}
  \caption{\emph{a}. Upper branch: Magnitude of the largest ($\alpha$) and second largest ($\beta$) components of a collection of $\bm{\pi}$-vectors sampled uniformly over entropy level sets for simplexes of different sizes. The sampling procedure is detailed in Appendix \ref{sec:simul}. Different colours represent different values of $K$. Solid lines: medians. Shaded areas: $99\%$ of the sampled values. For $H_0 > H_c$ (right end of the $x$-axis), all components are microscopic. For $H_0 < H_c$, the largest component $\alpha$ emerges as a macroscopic condensate. Lower branch: Second-largest sampled component $\beta$ scaled by the mass of the background fluid $(1 - \alpha)$. The second component remains negligible across the entire regime.  Solid gray line: Limiting behaviour of $\hat{\alpha}_{\mathrm{approx}} $ for $K \to +\infty$ (Eq.~\ref{eq:a_infty}).
  \emph{b}: Value of $\alpha$ obtained in different ways for $K = 500$. Green solid line: Sampled value (same as panel \emph{a}). Black dashed line: relation $\hat{\alpha}_{\mathrm{SP}}(H_0)$ obtained from solving the saddle-point equations Eq.~\eqref{app:eq:saddle-point} from Appendix \ref{sec:app:mean}) numerically. The match with the sampled values is remarkably precise. Gray dashed line: relation  $\hat{\alpha}_{\mathrm{approx}}(H_0)$ obtained by numerically solving Eq.~\eqref{eq:a_eq}, valid for $H_0 < H_c$. Solid gray line: Limiting behaviour of $\hat{\alpha}_{\mathrm{approx}} $ for $K \to +\infty$ (Eq.~\ref{eq:a_infty}).}   
  \label{fig:samples}
\end{figure}

In Appendix \ref{sec:app:mean}, we provide an explicit expression for the saddle-point equations (Eq.~\ref{app:eq:saddle-point}), which can be solved numerically to obtain $\alpha$ as a function of $H_0$ for a given $K$. In Fig.~\ref{fig:samples}\emph{b} (black dashed line), this theoretical prediction ($\hat{\alpha}_\mathrm{SP}$) is contrasted with the one obtained by sampling the entropy level surface of the simplex uniformly. The sampling procedure is detailed in Appendix \ref{sec:simul}.  The two approaches are in excellent agreement.

A simpler relation between $\alpha$ and $H_0$ can be obtained by exploiting the concentration of the entropy $H(\boldsymbol{\tilde\pi})$ of the fluid components around $H_c$. For $H_0 < H_c$, replacing $H(\boldsymbol{\tilde\pi})$ by $H_c$ in Eq.~\eqref{eq:nesting}, yields, up to $\mathcal{O}(1/\log K)$,
\begin{equation} \label{eq:a_eq}
  H_0 \approx H(\alpha, 1-\alpha) + (1-\alpha) \ H_c,
\end{equation}
Given $H_0$ and $K$, Eq.~\eqref{eq:a_eq} can be solved numerically for $\alpha$, yielding the $\hat\alpha_{\mathrm{approx}}$ displayed in Fig.~\ref{fig:samples}\emph{b} in gray dashed line. It provides a very good approximation to the size of $\alpha$ (note the logarithmic scale), though not as good as $\hat{\alpha}_{SP}$.

Dividing Eq.~\eqref{eq:a_eq} by $H_c$ and taking the limit $K \to \infty$ while keeping $H_0/H_c$ fixed yields the linear relation
\begin{equation} \label{eq:a_infty}
    \alpha = 1 - \frac{H_0}{H_c}
\end{equation}
shown by the straight black line of Fig.\ref{fig:samples}\emph{a}. This limiting behaviour is approached slowly, since the error in Eq.~\eqref{eq:a_eq} is of order $\mathcal{O}(1/\log K)$.

\subsection{The ranking of the fluid components below the critical entropy}

When $H_0 \le H_c$, a single condensate component acquires a macroscopic value $\alpha$. Figure~\ref{fig:samples}\emph{b} shows that the second largest component, of  magnitude $\beta$, remains microscopic, even after normalisation. In Sect.~\ref{sec:suppression}, we explain why the number of $\bm{\pi}$-vectors with more than a single condensate is negligible among those compatible with the entropy constraint $H = H_0$. We then here focus on $\bm{\pi}$-vectors of a single condensate component, and derive the ranking of all the $K-1$ remaining
fluid components. 

For small $\tilde{\pi}$-values, the single-site density $\tilde f(\tilde{\pi})$ of the normalised fluid components is well approximated by an exponential function, namely
\begin{equation*}
    \tilde f(\tilde{\pi}) \sim  K \mathrm{e}^{-K\tilde{\pi}} \ \ \ \text{for small }\tilde{\pi}. 
\end{equation*}
This approximation loses validity for $\tilde{\pi} \gg \sfrac{1}{K}$, but such values have a vanishing probability anyhow. The associated cumulative distribution is $F(\pi) = 1-\mathrm{e}^{-K \pi}$. Therefore, Eq.~\eqref{eq:rank} can be inverted, yielding the ranking of Eq.~\eqref{eq:triangulito}. This approximation is valid for all $H_0$ values equal or below $H_c$, as long as $K$ is large. The inset of Fig.~\ref{fig:h_samples} confirms this result for two different $H_0$. The rescaled fluid components of two vectors sampled from the corresponding level sets essentially overlap, and are very well described by the law of Eq.~\eqref{eq:triangulito}. As a consequence, we conclude that in level sets of $H_0 \le H_c$, the fluid components of almost all sampled vectors follow a universal ranking, which also coincides with the ranking found when sampling the whole simplex with uniform measure. 

\subsection{Suppression of multiple condensates}

\label{sec:suppression}

In principle, low $H_0$ values could be achieved by distributions $\bm{\pi}$
with more than one condensed component. Yet, the overwhelming majority of
$\bm{\pi}$-vectors actually sampled contain a single condensate. The second
largest component $\beta$ remains typically vanishingly small compared with
the largest component $\alpha$, as seen in the lower branch of
Fig.~\ref{fig:samples}\emph{a}. Thus, the system does not undergo a secondary
condensation within the fluid.

To understand why configurations with multiple condensates are so rare,
consider a configuration $\bm{\pi}^r$ that satisfies the entropy constraint
$H=H_0<H_c$ by distributing a macroscopic fraction of the total probability
among $r>1$ condensate components. Because the Shannon entropy is concave,
fragmenting a given condensate among several components increases the
entropy of the condensed phase. Thus, the entropy contribution of the
condensates in $\bm{\pi}^r$ is larger than that of a configuration
$\bm{\pi}^1$ with a single condensate whose magnitude equals the combined
weight of the $r$ condensates in $\bm{\pi}^r$. To satisfy the same entropy
constraint, $\bm{\pi}^r$ must drain additional probability from the
background fluid.

This comes at a substantial entropic cost. For a fixed condensate magnitude
$\alpha$, the fluid configurations are dominated by those in which the
renormalised fluid components have an exponentially decreasing rank and
entropy close to $H_c$, just as the full simplex is dominated by configurations
with exponentially decreasing rank and entropy close to $H_c$. Depleting the
fluid to sustain multiple condensates therefore reduces its capacity to
generate microscopic configurations. The resulting multiplicity of
multi-condensate states is exponentially smaller than that of single-condensate
states. Consequently, the system overwhelmingly favours a single condensate,
which sequesters the probability required by the macroscopic constraint while
leaving the largest possible mass in the background fluid.

\section{Conclusions}

\label{sec:discussion}

In Fig.~\ref{fig:samplesmicro} we displayed examples of $\bm{\pi}$-vectors sampled uniformly from level sets of entropy. The empirical distributions of components of the set of all such vectors are varied. Yet, in the limit of high dimensionality $K$, most of those distributions occupy a vanishing volume of the simplex. The analysis here presented demonstrated that each level set has a typical empirical distribution of components (or ranking), the shape of which depends on the chosen level set. 

When $H_0 = \log K$, all $\pi_i$ components have exactly the same value $\sfrac{1}{K}$. As $H_0$ starts to diminish, the components $\pi_i$ become more heterogeneous. In this regime, the ranking of components of the actually sampled vectors can be derived from the cumulative distribution associated with the single-site density $f(\pi)$ of Eq.~\eqref{eq:single_site_dist}. 

A phase transition takes place when $H_0$ reaches the critical entropy $H_c$ given by Eq.~\eqref{eq:Hc}, which is approximately equal to $H^\triangle = \log K + \gamma - 1$, that is, the entropy maximally represented in the entire simplex.
The homogeneous fluid cannot continue reducing its entropy through smooth and uniform adjustments of its microscopic components, so the system is forced to undergo condensation to satisfy the global constraints.  In this regime, the largest component $\alpha$ of the actually sampled $\bm{\pi}$-vectors detaches from the microscopic bulk and begins to absorb a macroscopic fraction of the total probability, growing steadily toward unity as $H_0 \to 0$, as shown in Fig.~\ref{fig:samplesmicro}. The numerical solution of Eq.~\eqref{eq:a_eq} shows that $\alpha$ increases continuously from zero as $H_0$ decreases below $H_c$, with no discontinuity at $H_c$. Thus, the condensate fraction $\alpha$ acts as a continuous order parameter, and the transition is \emph{second order}. The normalisation condition implies that the remaining components must remain, on average, below $\sfrac{1}{K}$.  The accumulation of probability in a single component is an effective way of reducing the entropy of the $\pi$-vector. Geometrically, distributions $\bm{\pi}$ with a condensed component belong to the lobes of the simplex.

Below the critical entropy, the empirical distribution of the normalised fluid components is governed by $\tilde{f}(\tilde{\pi})$, which is well approximated by an exponential. As a consequence, the ranked distribution of renormalised fluid components is governed by Eq.~\eqref{eq:triangulito} (Fig.~\ref{fig:samples}), that is, the distribution describing $\bm{\pi}$ vectors sampled uniformly from the whole unconstrained simplex. This suggests, that the fluid subsystem behaves almost independently of the condensate.



\appendix

\section{Distribution of components in the microcanonical ensemble}

\label{sec:pelotitas}

To quantify the fraction of $\bm{\pi}$ vectors with a given distribution of components, we introduce a discretization strategy. The total probability $\sum_j \pi_j = 1$ is viewed as a conserved resource partitioned into $S$ discrete probability quanta across $K$ categories. A specific configuration is defined by the vector $\boldsymbol{s}=(s_1,\dots,s_K)$, where $s_i$ represents the number of quanta in category $i$, subject to the constraint $\sum_{i=1}^K s_i=S$. The corresponding probability vector is given by $\pi_i=s_i/S$. In the microcanonical perspective, every distinct distribution of quanta is equally likely \cite{jaynes1957information, landau2013statistical}.

A macrostate $\boldsymbol{s}^*$ is defined as the set of all configurations $\boldsymbol{s}$ with the same set of occupancy values, regardless of their category assignments. To exhibit a macrostate, we display a representative of the set, namely, the one with components ranked in decreasing order. The number  $N(\boldsymbol{s}^*)$ of $\boldsymbol{s}$ states belonging to a given macrostate $\boldsymbol{s}^*$ is
\begin{equation}
    \displaystyle N(\boldsymbol{s}^*) = \frac{K!}{\prod_{s = 0}^S m_{s}!},
    \label{eq:pmacro}
\end{equation}
\noindent where $m_s=m_s(\boldsymbol{s}^*)$ denotes the number of categories (multiplicity) containing exactly $s$ quanta. By construction, the numbers $m_s$ associated with any macrostate $\boldsymbol{s}^*$ must satisfy the normalization constraint $\sum_{s=0}^S m_s=K$, and the mean occupancy constraint $\sum_{s=0}^S s \ m_s= S$.

We consider the thermodynamic limit where $S$ and $K$ are large, but the number of probability quanta per category $S/K$ is fixed. In this regime, the observed multiplicities are macroscopically large, so Stirling's approximation turns Eq.~\eqref{eq:pmacro} into
\begin{equation}
    \displaystyle N(\boldsymbol{s}^*) \propto \exp\left[K H\big(\boldsymbol{\hat{f}}(\boldsymbol{s}^*)\big)\right],
    \label{eq:pmacroHf}
\end{equation}
where $H(\boldsymbol{g})=-\sum_{j=0}^S g_j \log g_j$ is the entropy of the multiplicity distribution $\hat{f}
_s(\boldsymbol{s}^*) = m_s(\boldsymbol{s}^*)/K$. The component $\hat{f}_s$ measures the probability of observing a category with $s$ quanta in a given macrostate $\boldsymbol{s}^*$. Since in the microcanonical ensamble, the probability of a macrostate is proportional to the number of states it contains, Eq.~\eqref{eq:pmacroHf} implies that the probability of a macrostate scales exponentially with the entropy of its 
multiplicities. 

\section{Approximation for the single-site integral} \label{sec:single_site_int}

In order to later (in the following sections) assess the validity of the saddle-point approximation, in this section, we derive an asymptotic approximation for the single-site integral
\begin{align*}
  z(s,t)
  = \int_0^1 \mathrm{d}\pi \ \mathrm{e}^{-s\pi+t\pi\log\pi}, \qquad s,t > 0,
\end{align*}
in terms of inverse powers of $L(s,t) = \log t + s/t$, valid to order $\mathcal{O}(L(s,t)^{-2})$.
The approximation becomes exact in the limit $K \to +\infty$ with constant $H_0$, in which $s \to \infty$ and $t$ is bounded away from $0$, which also sends $L(s,t) \to \infty$. Since $t > 0$, the integrand exhibits a single minimum at $\pi_m = \mathrm{e}^{s/t-1}$. We can therefore approximate the integral by truncating it near this minimum,
\begin{align*}
    z(s,t)
    \approx \int_0^{\mathrm{e}^{s/t-1}} {\rm d}\pi \ \mathrm{e}^{-s \pi + t \pi \log \pi}.
\end{align*}
If $\pi_m < 1$, the tail contribution from $(\pi_m, 1)$ is discarded; if $\pi_m > 1$, the artificial integration domain $(1, \pi_m)$ is added. Because $s \to \infty$, the error introduced by these domain modifications decays exponentially with $K$.

Defining the function $L(s, t) = \log t + s/t$ and introducing the change of variables $x = t L(s,t) \pi$, the integral becomes
\begin{widetext}
\begin{align*}
    z(s,t)
    &= \frac{1}{t L(s,t)} \int_0^{t L(s,t)e^{s/t-1}} \mathrm{d}x \, \mathrm{e}^{-x} \exp\left(\frac{x \log x - x \log(L(s,t))}{L(s,t)}\right) \\
    &\approx \frac{1}{t L(s,t)} \int_0^{\infty} \mathrm{d}x \, \mathrm{e}^{-x} \Bigg[1 + \frac{x \log x - x \log(L(s,t))}{L(s,t)} + \frac{1}{2} \left(\frac{x \log x - x \log(L(s,t))}{L(s,t)}\right)^2\Bigg] \\
    &= \frac{1}{t L(s,t)} \Bigg[1 - \frac{\log(L(s,t)) + \gamma - 1}{L(s,t)} + \frac{\log(L(s,t))^2 - (3 - 2 \gamma) \log(L(s,t)) + \gamma^2 - 3\gamma + 1 + \frac{\pi^2}{6}}{L(s,t)^2}\Bigg],
\end{align*}
\end{widetext}
where $\gamma \approx 0.5772$ is the Euler-Mascheroni constant. Taking the logarithm, 
\begin{align}
    \log z(s,t)
    \approx & - \log t - \log(L(s,t)) - \label{app:eq:logz_approx} \\
    & \frac{\log(L(s,t)) + \gamma - 1}{L(s,t)} + \nonumber \\
    & \frac{\frac{1}{2} \log(L(s,t))^2 - (2-\gamma) \log(L(s,t)) + C}{L(s,t)^2},
    \nonumber
\end{align}
where $C = \frac{1}{2} \gamma^2 - 2\gamma + \frac{1}{2} + \frac{\pi^2}{6} \approx 1.1571$.

\section{Condensation threshold}
\label{sec:condensation_threshold}

Starting from ~\eqref{app:eq:logz_approx}, we here evaluate the saddle-point equations \eqref{eq:saddle-point} along the diagonal $t = s$ and show that, for this particular choice, the critial entropy is given by Eq.~\eqref{eq:Hc}.

We first write $\log z(s,t) = -\log t + g(L)$, with $L = L(s,t) = \log t + s/t$ and
\begin{align*}
    g(L) =& -\log L - \frac{\log L + \gamma - 1}{L} + \\ & \frac{\tfrac12\log(L)^2 - (2-\gamma)\log L + C}{L^2},
\end{align*}
where $C = \tfrac12\gamma^2 - 2\gamma + \tfrac12 + \tfrac{\pi^2}{6}$. Since $\partial_t L = 1/t - s/t^2$, the derivative vanishes along the diagonal,
\begin{align*}
    \partial_t L\big|_{t=s} = 0.
\end{align*}
The second saddle-point equation then simplifies to
\begin{equation*}
    \partial_t \log z\big|_{t=s} = -\frac1t + g'(L)\,\partial_t L\big|_{t=s} = -\frac{1}{s} = -\frac{H_c}{K},
\end{equation*}
implying that
\begin{equation} \label{eq:sn1}
    H_c = \frac{K}{s}.
\end{equation}
In turn, the first equation reads
\begin{equation} \label{eq:sn2}
    \partial_s \log z\big|_{t=s} = \frac{g'(L)}{s} = -\frac{1}{K}
    \quad\Longrightarrow\quad
    s = -K\,g'(L).
\end{equation}
Combining the two results in $H_c = -1/g'(L)$. Differentiating $g$ term by term,
\begin{align*}
    g'(L) =& -\frac{1}{L} + \frac{\log L + \gamma - 2}{L^2} +  \\
    & + \frac{-\log(L)^2 + (5-2\gamma)\log L + (\gamma - 2 - 2C)}{L^3},
\end{align*}
and expanding the reciprocal $-1/g'(L)$ in inverse powers of $L$, the $\log(L)^2$ contributions cancel and yield
\begin{align}
    H_c = & -\frac{1}{g'(L)} = L + \log L + \gamma - 2 \label{app:eq:H0_of_L} \\
    & + \frac{\log L + \gamma + 1 - \pi^2/3}{L} + \mathcal{O}\!\left(\frac{\log(L)^2}{L^2}\right).
    \nonumber
\end{align}
It remains to eliminate $L$ in favor of $K$. On the diagonal $L = \log s + 1$, and since $s = K/H_c$,
\begin{align}
    L = \log K - \log H_c + 1.
    \label{app:eq:L_of_K}
\end{align}
From~\eqref{app:eq:H0_of_L} we have $\log H_c = \log L + (\log L + \gamma - 2)/L + \mathcal{O}(\log(L)^2 / L^2)$. Inserting this into~\eqref{app:eq:L_of_K} and back into~\eqref{app:eq:H0_of_L}, every $\log L$ term cancels, leaving
\begin{align*}
    H_c = \log K + \gamma - 1 - \frac{\pi^2 - 9}{3\,L}.
\end{align*}
Finally, on the diagonal $L = \log K - \log H_c + 1 = \log K + \mathcal{O}(\log\log K)$. Replacing $L$ by $\log K$ in the $\mathcal{O}(1/L)$ correction introduces only an $\mathcal{O}(\log\log K / \log(K)^2)$ error, smaller than the terms already discarded, and recovers Eq.~\eqref{eq:Hc}.

\section{Failure of the symmetric saddle point in the condensed regime}

\label{sec:failure}

In this section we demonstrate that the symmetric saddle-point treatment of $Z(H_0)$ fails for $H_0 < H_c$. The argument proceeds in three steps. First, we isolate the single approximation on which the symmetric treatment rests, namely, the assumption that the values of $M$ and $H$ derived from $K$ independent samples of $f(\pi)$ (as prescribed by the canonical ensemble) is governed by a Gaussian distribution centred at the values prescribed by the constraints. Second, we decompose the single-site density into its fluid and condensed modes and extract the two properties of the condensed mode that drive the argument: that a condensed component carries (1) nearly a full unit of mass, and (2) essentially no entropy. Third, we compute the characteristic function of the joint probability density governing the pair $(M, H)$ and anti-transform
it explicitly, showing that in the condensed regime, the distribution is not a Gaussian, but a Poisson-weighted mixture of narrow Gaussian peaks. 


Throughout, $f(\pi) = z(s,t)^{-1} e^{-s\pi + t\pi\log\pi}$ denotes the single-site density of the main text, with $(s,t)$ fixed by the saddle-point equations.



\subsection{Structure of the density below threshold}

For $H_0 < H_c$ the exponent $\varphi(\pi) = -s\pi + t\pi\log\pi$ satisfies $t > s$.
The density $f$ is therefore bimodal, with a fluid mode supported near the origin and a condensed mode at $\pi \approx 1$, separated by the minimum at $\pi_c = e^{s/t - 1}$.
We decompose $f$ as the mixture of the two modes,
\begin{equation} \label{eq:mixture}
  f(\pi) = (1 - w_1)\, f_0(\pi) + w_1\, f_1(\pi),
\end{equation}
where
\begin{equation*}
  \left\{
  \begin{aligned}
    f_0(\pi) = &\frac{1}{z_0(s,t)}\, e^{-s\pi+t\pi\log\pi}\, \boldsymbol{1}_{[0,\pi_c]}(\pi), \ \ \mathrm{with} \\ &z_0(s,t) = \int_0^{\pi_c} {\rm d}\pi \ e^{-s\pi+t\pi\log\pi}, \\
    f_1(\pi) = & \frac{1}{z_1(s,t)}\, e^{-s\pi+t\pi\log\pi}\, \boldsymbol{1}_{(\pi_c,1]}(\pi), \ \ \mathrm{with} \\ & z_1(s,t) = \int_{\pi_c}^1 {\rm d}\pi \ e^{-s\pi+t\pi\log\pi},
  \end{aligned}
  \right.
\end{equation*}
the indicator function is $\boldsymbol{1}_A(x) = 1$ if $x \in A$ and zero otherwise, and $w_1 = z_1/z$ is the weight of the condensed mode. Under this decomposition, a component drawn from $f$ falls in the condensed mode with probability $w_1$ and in the fluid mode with probability $1 - w_1$, and every moment splits accordingly, $\langle \cdot \rangle_f = (1-w_1) \langle \cdot \rangle_{f_0} + w_1 \langle \cdot \rangle_{f_1}$.

Two properties of the condensed mode drive the entire argument.
Writing $\pi = 1 - v$ near the endpoint, the exponent is $\varphi(1-v) = -s + (t-s)v + \mathcal{O}(t v^2)$, so $f_1$ is exponentially concentrated on the scale $v \sim 1/(t-s)$, which vanishes as $K \to \infty$ since $t - s \to \infty$ below threshold.
Consequently a component drawn from $f_1$ carries
\begin{equation} \label{eq:condensed_content}
  \pi = 1 - \mathcal{O}\big(\tfrac{1}{t-s}\big), \qquad h(\pi) = \mathcal{O}\big(\tfrac{1}{t-s}\big),
\end{equation}
that is, nearly the entire mass budget and essentially no entropy, since $h(\pi) \to 0$ as $\pi \to 1$.

The first saddle-point equation bounds the condensed weight $w_1$. Since both $\langle \pi \rangle_{f_0}$ and $\langle \pi \rangle_{f_1}$ are non-negative, $w_1 \langle \pi \rangle_{f_1} \le \langle \pi \rangle_f = 1/K$, and $\langle \pi \rangle_{f_1} \approx 1$ by \eqref{eq:condensed_content}, so that
\begin{equation*}
  w_1 \lesssim \frac{1}{K}.
\end{equation*}
It is therefore natural to measure the condensed weight in units of $1/K$ and define
\begin{equation*}
  c = K w_1, \qquad 0 \le c \lesssim 1.
\end{equation*}
The value of $\langle \pi \rangle_{f_0}$ then follows from the mixture decomposition of $\langle \pi \rangle_f$,
\begin{equation} \label{eq:fluid_mass}
  \langle \pi \rangle_{f_0} = \frac{1}{1-w_1} \left( \frac{1}{K} - w_1 \langle \pi \rangle_{f_1} \right) \approx \frac{1 - c}{K},
\end{equation}
so that, summed over the $K$ components, the fluid mode carries a mean mass $1 - c$ while the condensed mode is expected to carry a mean mass $c$. 

Equation \eqref{eq:fluid_mass} also fixes the scale of the fluid components, which are of size $\pi = \mathcal{O}(1/K)$ up to logarithmic factors. To see this, rescale $\pi = u/K$ in the fluid exponent,
\begin{equation*}
  \varphi\big(u/K\big) = -\theta\, u + \frac{t}{K}\, u \log u, \qquad \theta = \frac{s + t \log K}{K}.
\end{equation*}
A mean of order $1/K$ requires $\theta = \Theta(1)$, and since $s > 0$ this forces $t \log K \le \theta K$, that is, $t/K = \mathcal{O}(1/\log K)$. The fluid mode is therefore approximately exponential on the scale $\pi \sim 1/K$, with corrections of relative order $1/\log K$ coming from the $u \log u$ term. In particular, its second moment is $\langle \pi^2 \rangle_{f_0} = \mathcal{O}(1/K^2)$ up to logarithmic factors, which we record as
\begin{equation} \label{eq:fluid_second}
  K \big\langle \pi^2 \big\rangle_{f_0} = \mathcal{O}\!\left( \frac{\log(K)^2}{K} \right).
\end{equation}
Similarly, since the condensed mode carries essentially no entropy by \eqref{eq:condensed_content}, the entropy constraint is met by the fluid mode alone, and the mixture decomposition of $\langle h \rangle_f = H_0/K$ gives
\begin{equation} \label{eq:fluid_entropy}
  \langle h(\pi) \rangle_{f_0} \approx \frac{H_0}{K}.
\end{equation}

\subsection{The limit law is a mixture of Gaussians}
\label{sec:mixture}

We now compute the limit of the joint characteristic function of the constrained pair,
\begin{equation*}
  \phi(k_1, k_2) = \Big\langle e^{\,i k_1 M + i k_2 H} \Big\rangle_{\tilde p} = \Big\langle e^{\,i k_1 \pi + i k_2 h(\pi)} \Big\rangle_f^{K},
\end{equation*}
where the factorization uses the independence of the components under the tilted product measure. The mixture decomposition \eqref{eq:mixture} splits the single-site expectation over the two modes,
\begin{align*}
  \Big\langle e^{\,i k_1 \pi + i k_2 h(\pi)} \Big\rangle_f =& (1 - w_1) \Big\langle e^{\,i k_1 \pi + i k_2 h(\pi)} \Big\rangle_{f_0} + \\
  & + w_1 \Big\langle e^{\,i k_1 \pi + i k_2 h(\pi)} \Big\rangle_{f_1},
\end{align*}
and the two terms behave differently. In the condensed mode, $k_1 \pi + k_2 h(\pi) = k_1 + \mathcal{O}\big((|k_1| + |k_2|)/(t-s)\big)$ by \eqref{eq:condensed_content}, so the exponential is $\mathcal{O}(1)$ and must be kept intact,
\begin{equation*}
  w_1 \Big\langle e^{\,i k_1 \pi + i k_2 h(\pi)} \Big\rangle_{f_1} = \frac{c}{K}\, e^{\,i k_1} \left( 1 + \mathcal{O}\Big(\tfrac{|k_1|+|k_2|}{t-s}\Big) \right).
\end{equation*}
In the fluid mode, both $k_1 \pi = \mathcal{O}(k_1/K)$ and $k_2 h(\pi) = \mathcal{O}(k_2 \log K / K)$ are small at fixed $(k_1, k_2)$, so the exponential may be expanded. Its first-order terms are the fluid moments \eqref{eq:fluid_mass} and \eqref{eq:fluid_entropy}, and its second-order terms involve the matrix $Q$ of second moments of $\big(\pi, h(\pi)\big)$ under $f_0$,
\begin{align*}
  (1 - w_1) \Big\langle e^{\,i k_1 \pi + i k_2 h(\pi)} \Big\rangle_{f_0} =& 1 - \frac{c}{K} + \frac{i k_1 (1-c)}{K} + \\
  & + \frac{i k_2 H_0}{K} - \tfrac{1}{2}\, \mathbf{k}^{\!\top} Q\, \mathbf{k} + \dots
\end{align*}
The entries of $Q$ are of order $\log(K)^2/K^2$, which follows from the fluid scale established above. Writing $\pi = u/K$ with $u$ of order one, the single-site entropy is $h(\pi) = \frac{u}{K}\big(\log K - \log u\big) = \mathcal{O}\big(\tfrac{\log K}{K}\big)$, so $\langle h(\pi)^2 \rangle_{f_0} = \mathcal{O}\big(\log(K)^2/K^2\big)$, while $\langle \pi^2 \rangle_{f_0} = \mathcal{O}(1/K^2)$ up to logarithmic factors by \eqref{eq:fluid_second}, and the cross moment is bounded by Cauchy--Schwarz. Adding the two modes and raising to the power $K$,
\begin{align} 
    \log \phi(k_1, k_2) =& i k_1 (1 - c) + i k_2 H_0 - \tfrac{1}{2}\, \mathbf{k}^{\!\top} C_0\, \mathbf{k} + 
    \nonumber\\ & + c \big( e^{\,i k_1} - 1 \big) + \mathcal{O}\!\left( \tfrac{\log K}{K} \right), \label{eq:levy}
\end{align}
where $C_0 = K Q = \mathcal{O}(\log(K)^2 / K)$ is the covariance of the fluid contributions to $(M, H)$, and the error holds at fixed $(k_1, k_2)$.

The density of the constrained pair is recovered from \eqref{eq:levy} by Fourier inversion,
\begin{equation*}
    \tilde p(m, h) = \frac{1}{(2\pi)^2} \int {\rm d}k_1 \int {\rm d}k_2 \ e^{-i k_1 m - i k_2 h}\ \phi(k_1, k_2),
\end{equation*}
and the inversion can be carried out explicitly. Expanding the factor contributed by the condensed mode,
\begin{equation*}
    e^{\,c \left( e^{i k_1} - 1 \right)} = \sum_{n \ge 0} \frac{e^{-c}\, c^n}{n!}\ e^{\,i k_1 n},
\end{equation*}
turns $\phi$ into a sum of terms whose exponents are linear and quadratic in $(k_1, k_2)$,
\begin{align*}
        \phi(k_1, k_2) \approx & \sum_{n \ge 0} \frac{e^{-c}\, c^n}{n!} \\
    & \exp\left[ i k_1 \big( (1-c) + n \big) +  i k_2 H_0 - \tfrac{1}{2}\, \mathbf{k}^{\!\top} C_0\, \mathbf{k} \right],
\end{align*}
and each term is a standard Gaussian integral. Term by term, the inversion gives
\begin{align} 
    \tilde p(m, h) & \approx \sum_{n \ge 0} \frac{e^{-c}\, c^n}{n!}\ \frac{1}{2\pi \sqrt{\det C_0}}\ \exp\left[ -\tfrac{1}{2}\, \mathbf{v}_n^{\!\top} C_0^{-1}\, \mathbf{v}_n \right], \nonumber \\
    \mathbf{v}_n & = \begin{pmatrix} m - (1-c) - n \\ h - H_0 \end{pmatrix}. \label{eq:comb}
\end{align}
 
The density is therefore a Poisson-weighted mixture of Gaussians, a sequence of narrow peaks of width $O(K^{-1/2})$ up to logarithmic factors, centered at mass values $(1-c) + n$ and all at entropy $H_0$. As $K \to \infty$ the peaks sharpen around their fixed centers while their weights remain Poisson, so \eqref{eq:comb} also identifies the limit law of the constrained pair,
\begin{equation} \label{eq:compound_poisson_sm}
    \big( M,\, H \big) \xrightarrow{\ d\ } \big( (1-c) + N,\ H_0 \big), \ \ \ N \sim \mathrm{Poisson}(c),
\end{equation}
a deterministic fluid contribution plus a Poisson number of components of unit mass, with no fluctuation surviving in the entropy direction at this order. The Poisson law has an elementary interpretation. Each component independently falls in the condensed mode with probability $w_1 = c/K$, so the number of condensed components is $\mathrm{Binomial}(K, c/K)$, which converges to $\mathrm{Poisson}(c)$, and each such component carries a unit of mass and no entropy by \eqref{eq:condensed_content}.
 
In the fluid regime, $c = 0$ and only the $n = 0$ term of \eqref{eq:comb} survives, so the mixture collapses to a single Gaussian peak centered at $(1, H_0)$, which is precisely the law on which the saddle-point evaluation \eqref{eq:gaussian_step} rests. For $c > 0$ the mixture is not a Gaussian. Each individual peak obeys a local central limit theorem, since it is built from the fluid components alone, but the peaks are separated by unit gaps in the mass direction, and the failure is that the constraint does not sit on any of them.


 
The density of the constrained pair is therefore a Poisson-weighted mixture of Gaussians, a sequence of narrow peaks of width $\mathcal{O}(K^{-1/2})$ up to logarithmic factors, centered at mass values $(1-c) + n$ and all at entropy $H_0$. In the fluid regime, $c = 0$ and only the $n = 0$ term survives, so the mixture collapses to a single Gaussian peak centred at $(1, H_0)$, which is precisely the law on which the saddle-point evaluation \eqref{eq:gaussian_step} rests. For $c > 0$ the mixture is not a Gaussian. Each individual peak obeys a local central limit theorem, since it is built from the fluid components alone, but the peaks are separated by unit gaps in the mass direction, and the failure is that the constraint does not sit on any of them.

This closes the argument. The exact identity \eqref{app:eq:saddle-point_app} requires $\tilde p$ at the point $(1, H_0)$. In the entropy direction this point sits at the center of every peak in \eqref{eq:comb}, consistent with the fact that the condensate carries no entropy. In the mass direction, since $0 < c < 1$,
\begin{equation*}
    1 - c \ < \ 1 \ < \ 2 - c,
\end{equation*}
so the constraint falls in the gap between the first two peaks, at a distance $\min(c,\, 1-c)$ from the nearest one. This distance exceeds the peak width by a factor of order $\sqrt{K}/\log K$. The tilted ensemble satisfies the mass constraint on average and violates it in every realization. The density $\tilde p(1, H_0)$ is therefore exponentially small in $K$, whereas the Gaussian evaluation \eqref{eq:gaussian_step} asserts a polynomially large value. The symmetric saddle-point treatment fails at this step, and the failure is confined to this step, since everything preceding the evaluation of $\tilde p$ in \eqref{app:eq:saddle-point_app} is exact.

\section{Mean and variance of the entropy of the normalized sub-components} \label{sec:mean_var_h}

Here, we analytically derive the asymptotic mean and variance of the entropy of the background components, conditional on the total macroscopic entropy of the system $H(\boldsymbol\pi) = H_0$.  The saddle-point equations characterizing the moments hold for any target entropy $H_0$. By contrast, their explicit closed-form evaluation is possible only in the condensed regime $H_0 < H_c \approx \log K + \gamma - 1$.

Let $\boldsymbol\pi = (\pi_1, \dots, \pi_K)$ be a probability vector, and let $\pi_{(i)}$ denote its components sorted in increasing order, such that
\begin{align*}
  \min\{\pi_1, \dots, \pi_K\} & = \pi_{(1)} \le \pi_{(2)} \le \dots \le \pi_{(K)} = \\
  & = \max\left\{\pi_1, \dots, \pi_K\right\}.
\end{align*}
The $K-1$ smallest components, normalized by their total probability mass $1-\pi_{(K)}$, form a valid probability vector on a $(K-2)$-dimensional sub-simplex.
We define these normalized sub-components such that $\tilde{\pi}_{(i)} = \pi_{(i)} / (1-\pi_{(K)})$ for $i = 1, \dots, K-1$.
Note that $\boldsymbol{\tilde \pi} = (\tilde{\pi}_1, \dots, \tilde{\pi}_{K-1})$ need not be sorted.

\subsection{Mean}
\label{sec:app:mean}

The partition function that generates the moments of the sub-component entropy is
\begin{align*}
  Z(H_0, b) =& \int_0^1 {\rm d}\pi_1 \dots \int_0^1 {\rm d}\pi_K \ \delta\left(\sum_{i=1}^K \pi_i - 1\right) \times \\
  & \times \delta\big(H(\boldsymbol\pi) - H_0\big) e^{-(K-1)b H(\boldsymbol{\tilde{\pi}})}.
\end{align*}

Because the integrand is invariant under permutations of the components of $\boldsymbol\pi$, we can simplify the domain of integration. By fixing one specific component to be the maximum (e.g., $\pi_K$), we can multiply the resulting integral by $K$, which accounts for the number of possible ``maximum candidates''. This yields,
\begin{align*}
  Z(H_0, b)
  =& K \int_0^1 {\rm d}\pi_K \int_0^{\pi_K} {\rm d}\pi_1 \dots \int_0^{\pi_K} {\rm d}\pi_{K-1} \times \\
  & \times \delta\left(\sum_{i=1}^K \pi_i - 1\right) \delta\big[H(\boldsymbol\pi) - H_0\big] e^{-(K-1) b H(\boldsymbol{\tilde{\pi}})}.
\end{align*}

To decouple the maximum component from the remaining ones, we introduce the change of variables $\alpha = \pi_K$ and $\tilde{\pi}_i = \pi_i/(1-\pi_K)$ for $i \in \{1, \dots, K-1\}$. Applying this transformation gives
\begin{align*}
   Z(H_0, b) =& 
  K \int_0^1 {\rm d}\alpha \ (1 - \alpha)^{K-2} \int_0^{\frac{\alpha}{1-\alpha}} {\rm d}\tilde{\pi}_1 \dots \int_0^{\frac{\alpha}{1-\alpha}} {\rm d}\tilde{\pi}_{K-1} \\
  &  \mathrm{e}^{-(K-1) b H(\boldsymbol{\tilde{\pi}})} \ \ \delta\left(\sum_{i=1}^{K-1} \tilde{\pi}_i - 1\right)   \\ &  \delta\big[H(\alpha, 1-\alpha) + (1-\alpha) H(\boldsymbol{\tilde{\pi}}) - H_0\big] ,
\end{align*}
where we have used the entropy decomposition property $H(\boldsymbol{\pi}) = H(\pi_{(K)}, 1 -\pi_{(K)}) + (1-\pi_{(K)}) H(\boldsymbol{\tilde{\pi}})$.

Next, we handle the constraints by using the exponential representation of the Dirac delta function,
\begin{align*}
  \delta(x) = \frac{K-1}{2 \pi i} \int_{-i \infty}^{i \infty} {\rm d}\tilde{s} \ e^{-(K-1)\tilde{s} x}.
\end{align*}
Substituting this representation for both constraints allows us to factorize the integrals over $\tilde{\pi}_i$:
\begin{widetext}
\begin{align*}
  Z(H_0, b)
  \propto& \int_{-i\infty}^{i\infty} {\rm d}\tilde{s} \int_{-i\infty}^{i\infty} {\rm d}\tilde{t} \ e^{(K-1)(\tilde{s}+\tilde{t} H_0)} \int_0^1 {\rm d}\alpha \ (1 - \alpha)^{K-2} e^{-(K-1)\tilde{t} H(\alpha, 1-\alpha)} \ \\
  & \int_0^{\frac{\alpha}{1-\alpha}} {\rm d}\tilde{\pi}_1 \dots {\rm d}\tilde{\pi}_{K-1} \ e^{-(K-1)\tilde s \sum_i \tilde{\pi}_i + (K-1)\left(b + \tilde t (1-\alpha)\right) \sum_i \tilde{\pi}_i \log \tilde{\pi}_i}  \\
  \propto& \int_{-i\infty}^{i\infty} {\rm d}\tilde{s} \int_{-i\infty}^{i\infty} {\rm d}\tilde{t} \ e^{(K-1)(\tilde{s}+\tilde{t} H_0)} \int_0^1 {\rm d}\alpha \ e^{(K-2) \log(1-\alpha) - (K-1)\tilde{t} H(\alpha, 1-\alpha)} \times \\
  & \times \left(\int_0^{\frac{\alpha}{1-\alpha}} {\rm d}\tilde{\pi} \ e^{(K-1)\left[-\tilde{s} \tilde{\pi} + \left(b + \tilde{t} (1-\alpha)\right) \tilde{\pi} \log \tilde{\pi}\right]}\right)^{K-1} \\
  \propto& \int_{-i\infty}^{i\infty} {\rm d}\tilde{s} \int_{-i\infty}^{i\infty} {\rm d}\tilde{t} \int_0^1 {\rm d}\alpha \ e^{-(K-1) g(\tilde{s}, \tilde{t}, \alpha, H_0, b)},
\end{align*}
\end{widetext}
where we define the exponent function as
\begin{align*}
  g(\tilde{s}, \tilde{t}, \alpha, H_0, b)
  &= -\tilde{s} - \tilde{t} H_0 - \left(1 - \frac{1}{K-1}\right) \log(1-\alpha) +  \\
  & \ \ \ \ +  \tilde{t} H(\alpha, 1-\alpha)  - \log \tilde{z}(\tilde{s}, \tilde{t}, \alpha, b) \\
  & \approx -\tilde{s} - \tilde{t} H_0 - \log(1-\alpha) + \\ 
 &  \ \ \ \  + \tilde{t} H(\alpha, 1-\alpha) - \log \tilde{z}(\tilde{s}, \tilde{t}, \alpha, b),
\end{align*}
and the single-site integral of the sub-components is given by
\begin{align*}
  \tilde{z}(\tilde s, \tilde t, \alpha, b) = \int_0^{\frac{\alpha}{1-\alpha}} {\rm d}\tilde{\pi} \ e^{(K-1)\left[-\tilde s \tilde{\pi} + \left(b + \tilde t (1-\alpha)\right) \tilde{\pi} \log \tilde{\pi}\right]}.
\end{align*}

To standardize the limits of this integral, we introduce a change of variables $u = \frac{1-\alpha}{\alpha}\tilde{\pi}$, which rewrites $\tilde z$ as
\begin{align*}
  \tilde{z} = \frac{\alpha}{1-\alpha} \int_0^1 {\rm d}u \ e^{(K-1)(\mu u + \sigma u \log u)},
\end{align*}
with the newly defined parameters:
\begin{align*}
  \mu &= \frac{\alpha}{1-\alpha}\left[-\tilde s + \big(b + \tilde t(1-\alpha)\big) \log\left(\frac{\alpha}{1-\alpha}\right)\right], \\
  \sigma &= \frac{\alpha}{1-\alpha} b + \alpha \tilde t.
\end{align*}
Consequently, $g$ can be expressed entirely in terms of $(\mu, \sigma, \alpha, H_0, b)$:
\begin{align*}
  g(\mu, \sigma, \alpha, H_0, b)
  \approx  & \frac{1-\alpha}{\alpha} \mu - \frac{H_0 + \log\alpha}{\alpha}\sigma  \\ & + b \frac{H_0 - H(\alpha,1-\alpha)}{1-\alpha} - \\ & - \log\alpha - \log \mathcal{I}(\mu,\sigma),
\end{align*}
where the integral $\mathcal{I}$ is defined as
\begin{align*}
  \mathcal{I}(\mu,\sigma) = \int_0^1 {\rm d}u \ e^{(K-1)(\mu u + \sigma u \log u)}.
\end{align*}

Assuming large $K$, we approximate the partition function using the saddle-point method, obtaining
\begin{align*}
  Z(H_0,b) \appropto e^{-(K-1) g^*(H_0,b)}.
\end{align*}
Here, $g^*(H_0,b) = g\big(\mu^*(H_0, b), \sigma^*(H_0, b), \alpha^*(H_0,b), H_0, b\big)$, where the saddle-point values $\mu^*$, $\sigma^*$, and $\alpha^*$ are determined by the extremum conditions:
\begin{align*}
  \partial_\mu g =& \frac{1-\alpha^*}{\alpha^*} - \partial_\mu \log \mathcal{I}(\mu^*,\sigma^*) = 0, \\
  \partial_\sigma g =& -\frac{H_0 + \log \alpha^*}{\alpha^*} - \partial_\sigma \log \mathcal{I}(\mu^*, \sigma^*) = 0, \\
  \partial_\alpha g =& \frac{(H_0 + \log \alpha^* - 1)\sigma^* - \mu^* - \alpha^*}{(\alpha^*)^2} + \\
  & + b\frac{H_0 - H(\alpha^*,1-\alpha^*)-(1-\alpha^*)\log\frac{1-\alpha^*}{\alpha^*}}{(1-\alpha^*)^2} = 0.
\end{align*}

\begin{figure}[t]
    \centering
    \includegraphics[width=1\linewidth]{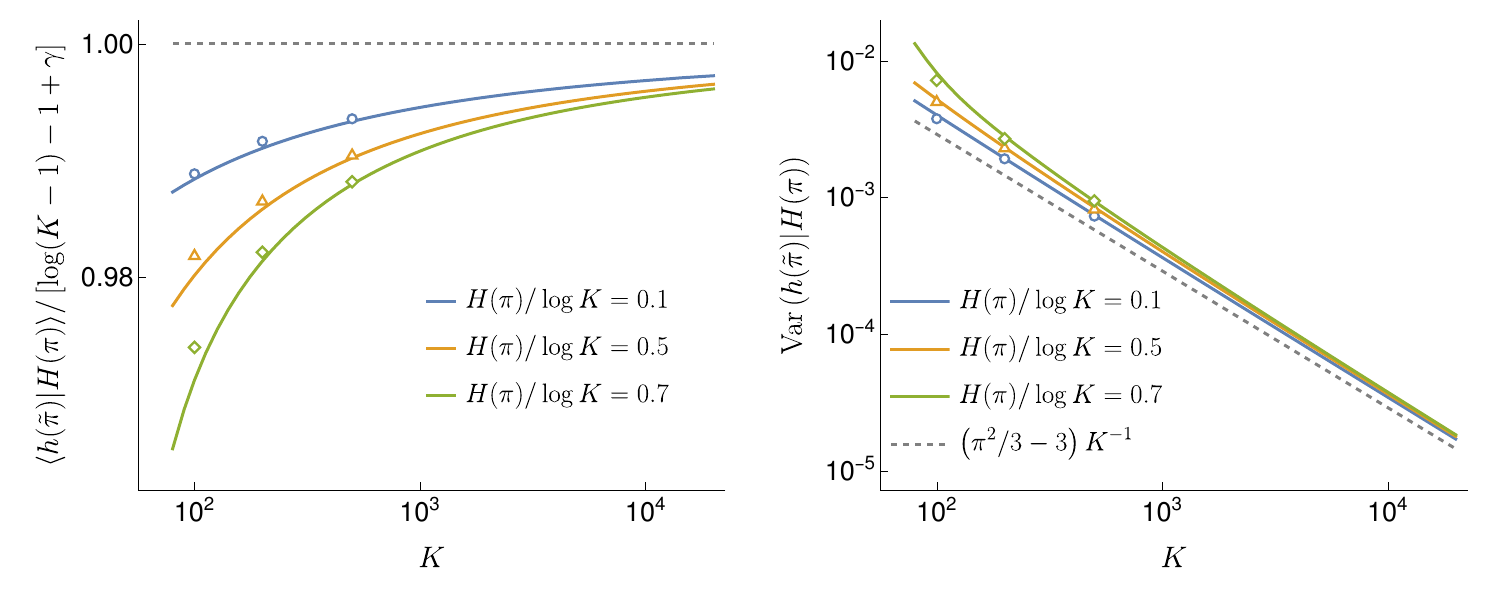}
    \caption{Conditional mean (left) and variance (right) of the normalized sub-component entropy, $H(\boldsymbol{\tilde{\pi}})$, as a function of the system size $K$. Solid lines represent analytical predictions derived from the exact saddle-point approximations (Eqs.~\eqref{app:eq:meanHx} and \eqref{eq:app:var_Hx}), while dots indicate empirical values computed from STAN Monte Carlo simulations. Colors correspond to fixed macroscopic entropy densities $H(\boldsymbol\pi)/\log K \in \{0.1, 0.5, 0.7\}$, all such that $H(\boldsymbol \pi) < H_c$. Left panel: To highlight asymptotic behavior, the conditional mean is scaled by its asymptotic limit, $\log(K-1) + \gamma - 1$, with the dashed grey line at $1.0$ indicating the convergence predicted by Eq.~\eqref{eq:app:asymp_mean_h}. Right panel: The unscaled conditional variance is shown, where the dashed grey line indicates the asymptotic variance predicted by Eq.~\eqref{eq:app:asymp_var}.}
    \label{fig:AppE:Mean_Var_vs_K}
\end{figure}

Finally, we can evaluate the mean entropy of the normalized sub-components by taking the derivative of the logarithm of the partition function with respect to $b$ at $b=0$:
\begin{align}
  \langle H(\boldsymbol{\tilde{\pi}}) | H(\boldsymbol\pi) = H_0 \rangle
  &= -\frac{1}{K-1} \left. \partial_b \log Z(H_0, b) \right|_{b=0} \nonumber \\
  &\approx \left. \partial_b g^*(H_0, b) \right|_{b=0} \nonumber \\
  &= \left. \partial_b g(\mu^*, \sigma^*, \alpha^*, H_0, b) \right|_{b=0} \nonumber \\
  &= \frac{H_0 - H(\alpha^*, 1-\alpha^*)}{1-\alpha^*}, \label{app:eq:meanHx}
\end{align}
where we used that $\partial_b g^*(H_0, b) = \partial_b g(\mu^*, \sigma^*, \alpha^*, H_0, b)$, since the implicit dependencies through the saddle-point parameters vanish ($\partial_\mu g = \partial_\sigma g = \partial_\alpha g = 0$).
At $b=0$, the saddle-point parameters $(\mu^*, \sigma^*, \alpha^*)$ satisfy the simplified system of equations:
\begin{align}
  \frac{1-\alpha^*}{\alpha^*} &= \partial_\mu \log \mathcal{I}(\mu^*,\sigma^*), \nonumber \\
  \frac{H_0 + \log \alpha^*}{\alpha^*} &= - \partial_\sigma \log \mathcal{I}(\mu^*,\sigma^*), \nonumber \\
  (H_0 + \log \alpha^* - 1)\sigma^* &= \mu^* + \alpha^*. \label{app:eq:saddle-point}
\end{align}
Eq. \eqref{app:eq:meanHx} and the saddle-point system \eqref{app:eq:saddle-point} make no assumption on $H_0$ and remain valid in both the fluid ($H_0 > H_c$) and condensed ($H_0 < H_c$) regimes. Solved numerically, they yield the solid curves shown in Fig.~\ref{fig:AppE:Mean_Var_vs_K}. The remainder of this section specializes to the condensed regime $H_0 < H_c$, in which $\mathcal{I}(\mu^*, \sigma^*)$ admits the explicit asymptotic treatment of Section~\ref{sec:single_site_int} and the moments acquire closed forms.

To this end, we approximate the integral $\mathcal{I}(\mu^*, \sigma^*)$ evaluated at the saddle point.
We first note that $\mathcal{I}(\mu,\sigma) = z\left(-(K-1)\mu,\, (K-1)\sigma\right)$, where $z$ (not $\tilde z$) is the single-site integral of Section~\ref{sec:single_site_int}. Guided by numerical solutions in the regime $H_0 < H_c$, we assume $\mu^* < 0$ and $\sigma^* > 0$, both bounded away from zero. This places the arguments of $z$ at $s = -(K-1)\mu^* = \mathcal{O}(K)$ and $t = (K-1)\sigma^* = \mathcal{O}(K)$, so approximation \eqref{app:eq:logz_approx} of Section~\ref{sec:single_site_int} applies. We obtain
\begin{align}
    \log \mathcal{I}(\mu,\sigma)
    \approx & -\log(K-1) - \log \sigma - \log(L(\mu,\sigma)) - \nonumber \\
    & - \frac{\log(L(\mu,\sigma)) + \gamma - 1}{L(\mu,\sigma)} + \frac{C}{L(\mu, \sigma)^2}\label{app:eq:logI_approx} \\
    &\quad + \frac{\frac{1}{2} \log(L(\mu,\sigma))^2 - (2-\gamma) \log(L(\mu,\sigma))}{L(\mu,\sigma)^2},
    \nonumber
\end{align}
where $L(\mu,\sigma) = \log(K-1) + \log \sigma - \mu/\sigma$ and $C = \frac{1}{2} \gamma^2 - 2\gamma + \frac{1}{2} + \frac{\pi^2}{6} \approx 1.1571$.

Plugging this expansion back into the saddle-point equations \eqref{app:eq:saddle-point} (a derivation analogous to the one in Section \ref{sec:condensation_threshold}), the mean entropy of the sub-components up to $\mathcal{O}(1/\log K)$ is
\begin{align}
    \langle H(\boldsymbol{\tilde{\pi}}) | H(\boldsymbol\pi) =  H_0 \rangle
    \approx & \log(K-1) + \gamma - 1 - \nonumber \\ & \frac{\pi^2 - 9}{3\log K}.
\label{eq:app:asymp_mean_h}
\end{align}
Provided $H(\boldsymbol{\pi}) < H_c$, Eq.~\eqref{eq:app:asymp_mean_h} confirms that in the limit of large $K$ the conditional mean converges to the unconstrained limit $\langle H(\boldsymbol{\tilde{\pi}}) | H(\boldsymbol\pi) \rangle \approx \log(K-1) + \gamma - 1$.

The left panel of Figure \ref{fig:AppE:Mean_Var_vs_K} visualizes this behavior, showing the conditional mean as a function of the system size $K$. The solid analytical curves, evaluated directly from the exact saddle-point formulation in Eq.~\eqref{app:eq:meanHx}, demonstrate excellent agreement with empirical simulations and beautifully capture the asymptotic convergence toward the unconstrained limit.

\subsection{Variance}

\label{sec:variance}

To compute the variance of the normalized sub-components, we utilize the generating properties of the partition function $Z(H_0,b)$. The variance is given by the second derivative of the log-partition function with respect to $b$, evaluated at $b=0$:
\begin{align*}
  \mathrm{Var}(H(\boldsymbol{\tilde{\pi}}) | H(\boldsymbol\pi) = H_0) = \frac{1}{(K-1)^2} \partial_b^2 \log Z(H_0,b)\Big|_{b=0}
\end{align*}

Applying the saddle-point approximation for large $K$, $Z(H_0,b) \appropto e^{-(K-1)g^*(H_0,b)}$, the variance simplifies to the second derivative of the saddle-point free energy:
\begin{align*}
  \mathrm{Var}(H(\boldsymbol{\tilde{\pi}}) | H(\boldsymbol\pi) = H_0) \approx -\frac{1}{K} \partial_b^2 g^*(H_0,b) \Big|_{b=0}
\end{align*}

To find the second derivative of the free-energy, we differentiate the first derivative $\partial_b g^* = \partial_b g$ with respect to $b$:
\begin{align*}
  \partial_b^2 g^* =  \partial_b \left(\partial_b g\right) = &\partial_b^2 g + (\partial_\mu \partial_b g) (\partial_b \mu^*) + \\ 
  & + (\partial_\sigma \partial_b g) (\partial_b \sigma^*) + (\partial_\alpha \partial_b g) (\partial_b \alpha^*)
\end{align*}
Because $b$ appears linearly in $g$ and only couples to $\alpha$, the terms $\partial_b^2 g$, $\partial_\mu \partial_b g$, and $\partial_\sigma \partial_b g$ are exactly zero. Expanding $H(\alpha^*, 1-\alpha^*)$, the only non-zero mixed partial derivative is:
\begin{align*}
  \partial_\alpha \partial_b g =& \partial_\alpha \left[ \frac{H_0 + \alpha^*\log\alpha^* + (1-\alpha^*)\log(1-\alpha^*)}{1-\alpha^*} \right] \\  =& \frac{H_0+\log\alpha^*}{(1-\alpha^*)^2}
\end{align*}
Thus, the second derivative simplifies to:
\begin{align*}
  \partial_b^2 g^* = \frac{H_0+\log\alpha^*}{(1-\alpha^*)^2} \partial_b \alpha^*.
\end{align*}

To find $\partial_b \alpha^*$, we apply the implicit function theorem to the extremum conditions. Let $\vec{v}^*(H_0, b) = (\mu^*(H_0,b), \sigma^*(H_0,b), \alpha^*(H_0,b))^T$. Taking the partial derivative of the gradient condition $\nabla_{\vec{v}} g = 0$ with respect to $b$ yields:
\begin{equation*}
  H_g \left(\partial_b \vec{v}^*\right) + \nabla_{\vec{v}}\left(\partial_b g\right) = 0  
\end{equation*}
so that
\begin{equation*}
    \partial_b \vec{v}^* = -H_g^{-1} \nabla_{\vec{v}}\left(\partial_b g\right)
\end{equation*}
where $H_g$ is the $3 \times 3$ Hessian matrix of $g$ (with respect to $\vec{v}$) evaluated at the saddle point.
Since $\partial_b g$ depends only on $\alpha$, its gradient vector is $\nabla_{\vec{v}} (\partial_b g) = \left(0, 0, \frac{H_0+\log\alpha^*}{(1-\alpha^*)^2}\right)^T$.
Extracting the third component of $\partial_b \vec{v}^*$ isolates the bottom-right element of the inverted Hessian matrix:
\begin{align*}
  \partial_b \alpha^* = - (H_g^{-1})_{33} \frac{H_0+\log\alpha^*}{(1-\alpha^*)^2}.
\end{align*}

By the block matrix inversion identity, $(H_g^{-1})_{33} = 1/S_\alpha$, where $S_\alpha$ is the Schur complement of the Hessian with respect to the variable $\alpha$, evaluated at the saddle-point $(\mu^*, \sigma^*, \alpha^*)$.

Substituting $\partial_b \alpha^*$ back into the variance equation yields the exact leading-order asymptotic behavior:
\begin{align}
  \mathrm{Var}(H(\boldsymbol{\tilde{\pi}}) | H(\boldsymbol\pi) = H_0)
  &\approx -\frac{1}{K} \left( \frac{H_0 + \log \alpha^*}{(1-\alpha^*)^2} \partial_b \alpha^* \right) \Bigg|_{b=0} \nonumber \\
  &= \frac{1}{K\,S_\alpha} \left[ \frac{H_0+\log\alpha^*}{(1-\alpha^*)^2} \right]^2,
  \label{eq:app:var_Hx}
\end{align}
where the quantities after the last equality should be evaluated at the solution to the saddle-point equations \eqref{app:eq:saddle-point}.

To evaluate the Schur complement $S_\alpha = (H_g)_{33} - (H_g)_{3,1:2} [(H_g)_{1:2,1:2}]^{-1} (H_g)_{1:2,3}$, we need to compute the second derivatives of $g$ with respect to $\mu$ and $\sigma$, which form the $2 \times 2$ sub-matrix $(H_g)_{1:2,1:2}$. These require the second derivatives of $\log \mathcal{I}(\mu, \sigma)$. Rather than differentiating the truncated asymptotic expansion, we compute these terms directly from the unexpanded integral $\mathcal{I}(\mu, \sigma)$ to preserve the positive-definiteness of the matrix.
Evaluating the relevant integrals by the same steps as in Section~\ref{sec:single_site_int},
we find that the determinant of $(H_g)_{1:2,1:2}$ is proportional to a constant $\Delta = \frac{\pi^2}{3} - 3$. Furthermore, provided $H_0 < H_c$, the saddle-point equations \eqref{app:eq:saddle-point} imply the relation $\mu^* \approx -\sigma^*$. This simplifies the mixed derivatives vector $(H_g)_{1:2,3}$, and combining these results yields the dominant term for the Schur complement:
\begin{align*}
    S_\alpha \approx \frac{L(\mu^*, \sigma^*)^2}{(\alpha^*)^2 (1-\alpha^*)^2 \Delta}.
\end{align*}

Plugging the expression for $S_\alpha$ into Equation \eqref{eq:app:var_Hx}, and solving the saddle-point equations \eqref{app:eq:saddle-point} with the approximation for $\log \mathcal{I}(\mu, \sigma)$ (Eq.~\eqref{app:eq:logI_approx}), the variance asymptotically is:
\begin{align}
    \mathrm{Var}(H(\boldsymbol{\tilde{\pi}}) | H(\boldsymbol\pi) = H_0)
    &\approx \frac{\pi^2 - 9}{3 \, K}.
    \label{eq:app:asymp_var}
\end{align}
This analytical result shows that the asymptotic conditional variance scales as $1/K$ and is independent of the macroscopic target entropy $H_0$.

The right panel of Figure \ref{fig:AppE:Mean_Var_vs_K} illustrates this conditional variance as a function of the system size $K$. The solid analytical curves, evaluated from the saddle-point formulation in Eq.~\eqref{eq:app:var_Hx}, show excellent agreement with the empirical variances computed via STAN simulations, holding remarkably well even for systems as small as $K = 100$.

Crucially, both the exact theoretical curves and the empirical data confirm that the variance decays as $1/K$ for any fixed entropy density $H(\boldsymbol\pi)/\log K$ such that $H(\boldsymbol\pi) < H_c$. This asymptotic decay demonstrates a concentration of measure: as $K$ grows, the conditional distribution of the sub-component entropy heavily concentrates around its mean value, which asymptotically is $\langle H(\boldsymbol{\tilde{\pi}}) | H(\boldsymbol\pi) \rangle \approx \log (K-1) + \gamma - 1$.


\section{Sampling procedure}
\label{sec:simul}

Our objective is to sample distributions $\pi$'s from different level curves of constant entropy $H(\pi)=H_0$ from a high-dimensional simplex $\Delta_{K-1}$. A naive approach would correspond to using a uniform prior $p(\pi)=\text{const}$ over the simplex. However, as we have mentioned in the main text, in this case most sampled distributions would have high entropy $H \simeq O(\log K)$ \citep{nemenman2001entropy}, as there is a concentration of measure towards $H(\pi)=H_c$ \citep{raginsky2013concentration}. In order to reduce this tendency of sampling high-entropy distributions, we propose to use a MaxEnt\citep{jaynes1957information, jaynes1982rationale, bialek2012biophysics} prior that has an exponentially smaller probability (in $K$) of sampling $\pi$'s with higher entropy. The exponential dependency is mediated by a hyperparameter $\lambda$ as
\begin{equation}
\displaystyle p(\pi|\lambda) =\frac{\exp\left[-\lambda K H(\pi)\right]}{Z(\lambda)} \propto \prod_{k=1}^K \pi_k^{-\lambda K\pi_k},
\label{eq:HMaxEnt}
\end{equation}
\noindent where $Z(\lambda)$ is the partition function that ensures normalization. Beyond being the least structured prior that has a constrained in entropy, a key property of the distribution from Eq.~(\ref{eq:HMaxEnt}) is that \emph{every probability vector $\pi$ sampled from an entropy level curve is equally likely}, which is not achievable with a Dirichlet prior \citep{nemenman2001entropy}. Then, changing $\lambda$, the dual coordinate to entropy \citep{amari2016information}, we can sample distributions from different parts of the simplex, and within each level curve of entropy analyze the properties of the sampled $\pi$'s.

This type of statistical model is also known as an exponential family \citep{amari2016information}. Such a prior is no longer conjugate to the multinomial, resulting in a loss of analytical tractability of the posterior evidence, a feature that we had with the Dirichlet distribution. However, we can still perform numerical sampling to obtain distributions from this prior. 

We perform numerical sampling through an adaptive Hamiltonian Monte Carlo (HMC) algorithm \citep{neal2011mcmc, betancourt2015hamiltonian, hoffman2014no} implemented in STAN \citep{Stan}, which can properly sample high-dimensional distributions $\pi$ that originate from our MaxEnt prior $p(\pi|\lambda) \propto \exp\left[-\lambda K H(\pi)\right]$. We generate hyperparameters in the range $\lambda \in (0,2)$. In the Monte Carlo algorithm, we generated $8$ chains, that have an initial warming phase of $1000$ steps, which were sufficient to observe the convergence of the mean entropy. For the next $2000$ steps, we took a sample every $10$ steps.




\bibliography{apssamp}

\end{document}